\documentclass[multphys,vecphys]{svmult}
\usepackage{makeidx}     
\usepackage{graphicx}    
\usepackage{multicol}    
\usepackage{latexsym}
\usepackage{bm}
\makeindex             

\newcommand{\tindex}[1]{\index{#1}#1}

\begin{document}

\title*{Exploring the nucleus in the context of low-energy QCD}
\author{Dario Vretenar\inst{1} and
Wolfram Weise\inst{2}}
\institute{Physics Department, Faculty of Science, University of
Zagreb, 10 000 Zagreb, Croatia
\texttt{[vretenar@phy.hr]}
\and ECT$^*$, I-38050 Villazzano (Trento), Italy and
Physik-Department, Technische Universit\"at M\"unchen,
D-85747 Garching, Germany \texttt{[weise@ect.it]}}
%
%
\maketitle
\abstract
These lecture notes address a central problem of theoretical nuclear physics: how to establish a relationship between {\em low-energy, non-perturbative QCD} and {\em nuclear phenomenology} which includes both nuclear matter and finite nuclei. We develop a microscopic covariant description of nuclear many-body dynamics constrained by chiral symmetry and in-medium QCD sum rules. A relativistic point-coupling model is derived, based on an effective Lagrangian with density-dependent contact interactions between nucleons. These interactions are constructed from chiral one- and two-pion exchange, combined with the large isoscalar nucleon self-energies that arise through changes in the quark condensate and the quark density at finite baryon density. Nuclear binding and saturation are almost completely generated by chiral (two-pion exchange) fluctuations in combination with Pauli
effects, whereas strong scalar and vector fields of about equal magnitude and opposite sign, induced by changes of the QCD vacuum in the presence of baryonic matter, generate the large effective spin-orbit potential in finite nuclei.  Promising results are found for the nuclear matter equation of state and for the bulk and single-nucleon properties of finite nuclei. 

\section{Introduction}
The physics based on Quantum Chromodynamics (QCD), the fundamental theory of
interacting quarks and gluons, is undoubtedly one of the most fascinating 
fields of modern science. It covers an enormously wide spectrum of phenomena, ranging from hadrons and nulcei to matter under extreme conditions of temperature and density. How the basic quark constituents and gluon fields arrange themselves to form nucleons and mesons, and how these  localized clusters of confined quarks and gluons interact collectively in nuclei: these are key questions of nuclear physics.

At very short distance scales, $r < 0.1$ fm,  corresponding to momentum transfers
above several GeV/c which probe space-time intervals deep inside the nucleon, QCD is a theory of weakly interacting, pointlike quarks and gluons. The rules for their dynamics are set by local gauge invariance under $SU(3)_{color}$. This is the area of perturbative QCD, an active field of research all by itself but with little direct impact on the understanding of nuclei in which the relevant length scale (e.g. the average separation between nucleons) is typically more than twice the size of the individual nucleon. At such large distance scales, $r > 1$ fm, QCD is realised as a theory of pions coupled to nucleons (and possibly other heavy, almost static hadrons). Their dynamics is governed by an approximate symmetry of QCD: {\it Chiral Symmetry}, and its spontaneous breakdown at low energy. 

The non-linear dynamics of gluons and their strong interactions with quarks imply a complex structure of the QCD ground state (or "vacuum"): it hosts strong condensates of quark-antiquark pairs and gluons, an entirely non-perturbative phenomenon. The quark condensate $\langle \bar{q}q \rangle$, i.e. the ground state expectation value of the scalar quark density, plays a particularly important role as an order parameter of spontaneously broken chiral symmetry. Hadrons, as well as nuclei, are excitations built on this condensed ground state. Changes of the condensate structure of the QCD vacuum in the presence of baryonic matter are a source of strong fields experienced by the nucleons. The nuclear dynamics governed by these strong fields, together with the pion-exchange forces that determine the nucleon-nucleon interaction at long and intermediate distances, will be important elements in our discussion.
 
Our framework is the effective field theory that represents low-energy QCD with the lightest ($u$ and $d$) quarks. This theory is written in terms of an effective Lagrangian constructed according to the symmetries and symmetry breaking patterns of QCD. Identifying the relevant, active degrees of freedom at low energy is the first important step in this approach. Nuclei are aggregates living in the low-temperature (hadronic) sector of the QCD phase diagram, where quarks and gluons are confined in color singlet mesons and baryons. At the energy and momentum scales characteristic of nuclei, only the states of lowest mass in the meson and baryon spectrum are expected to be "active": pions and nucleons. Their quark-gluon substructure and detailed short-distance dynamics are not resolved at nuclear scales. In the effective field theory the substructures of pions and nucleons are encoded through measured structure constants and through contact interactions with parameters to be determined empirically. The effective field theory approach does have predictive power, in the sense that there exists a guiding (symmetry) principle for constructing an ordered hierarchy of terms in the effective Lagrangian. The parameters of the leading terms are controlled or constrained by measurements of independent subprocesses such as pion-nucleon scattering.

Nuclear phenomenology has, of course, a long history and a rich empirical background that we can build on. Concepts such as density functional theory and effective field theory are at the basis of many successful nuclear structure models. Quantum Hadrodynamics (QHD), in particular, represents a class of Lorentz-covariant, meson-nucleon or point-coupling models of nuclear dynamics, that have been used successfully in studies of a variety of nuclear phenomena~\cite{SW.97}. Models based on the relativistic mean-field (RMF) approximation of QHD have been applied to describe various properties of spherical and deformed nuclei all over the periodic table~\cite{Rin.96}.

Pions, as Goldstone bosons of spontaneously broken chiral symmetry, are well known to be important ingredients in nuclear systems \cite{EW}. The behavior of the nucleon-nucleon interaction at long and intermediate distances is determined by one- and two-pion exchange processes. Ab-initio many-body calculations of nuclear matter and light nuclei \cite{Pan} strongly emphasize the crucial role played by the pion-exchange tensor force. Nevertheless, pions are usually not included as explicit degrees of freedom in QHD-type calculations.  The most successful QHD models are so far purely phenomenological. Effects of two-pion exchange are treated only implicitly through an ad-hoc scalar-isoscalar mean field. 

The successes of relativistic nuclear mean-field approaches have been attributed primarily to the large Lorentz scalar and four-vector nucleon self-energies which are at the basis of all QHD models \cite{FS.00}. There is substantial (though indirect) evidence, in particular from nuclear matter saturation and abnormally large spin-orbit splittings in nuclei, that the magnitude of each of the scalar and vector mean fields experienced by nucleons in the nuclear interior is of the order of several hundred MeV. Studies based on finite-density QCD sum rules have shown how such large scalar and vector nucleon self-energies can arise naturally from the in-medium changes of the scalar quark condensate and the quark density \cite{CFG.91,DL.90,Jin.94}. In most applications of QHD models to finite nuclei, however, the interactions generating these large nucleon self-energies have been treated in a purely phenomenological way. In some models they arise from the exchange of "sigma'' and "omega'' bosons with adjustable coupling strengths. The physical meaning of such assumed ad-hoc mechanisms may be seen in different perspective when exploring their basic QCD connection.  

These lecture notes describe and summarize recent developments based on the following conjectures that are believed to establish links between low-energy QCD, its symmetry breaking pattern, and the nuclear many-body problem: 
\begin{itemize} 
\item 
The nuclear ground state is characterized by large scalar and vector fields of approximately equal magnitude and opposite sign. These fields have their origin in the in-medium changes of the scalar quark condensate (the chiral condensate) and of the quark density.\\ 
\item 
Nuclear binding and saturation arise predominantly from chiral (pionic) fluctuations (reminiscent of van der Waals forces) in combination with Pauli blocking effects, superimposed on the condensate background fields and calculated according to the rules of in-medium chiral perturbation theory.\\
\end{itemize}
We begin by briefly summarizing important aspects of low-energy QCD and chiral symmetry inasmuch as they are relevant to our theme. Elements of in-medium chiral perturbation theory and results from QCD sum rules at finite density are also described. We then proceed by studying nuclear matter and nucleon self-energies in the context of chiral dynamics and with constraints from the expected changes of condensates at finite baryon density. In the next step, these features are translated into an equivalent relativistic point-coupling model with density dependent coupling strengths. This model is used to calculate a variety of bulk and single particle properties of finite nuclei, from $^{16}$O to $^{90}$Zr. We conclude with an outlook and a discussion of further steps that are still ahead.    

Much of the material and many of the results presented here have appeared in recent publications
\cite{KFW1,KFW2,KFW3,FKVW1,FKVW2}. A presentation of the general framework can be found in previous lecture notes \cite{Wei}. Introductory surveys on basic QCD notions related to our subject are given in several chapters of ref. \cite{TW}. 

\section{Low-energy QCD}

\subsection{Chiral symmetry}

Our framework is QCD in the sector of the lightest ($u$, $d$) quarks. They form a flavour $N_f = 2$ (isospin) doublet with "bare" quark masses of less than 10 MeV. The flavour (and colour) components of the quarks are collected in the Dirac fields $\psi (x) = (u(x), d(x))^T$. The \tindex{QCD Lagrangian},
\begin{equation}
{\cal L}_{QCD} = \bar{\psi} (i \gamma_{\mu} D^{\mu} - m) \psi - \frac{1}{2} Tr
(G_{\mu \nu} G^{\mu \nu}),
\label{QCD}
\end{equation}
involves the $SU(3)_{color}$ gauge covariant derivative $D^{\mu}$ and the gluonic field tensor $G^{\mu \nu} = (i/g) [ D^{\mu}, D^{\nu} ]$. The $2 \times 2$ matrix $m = diag (m_u, \, m_d)$ contains the light quark masses. The strange quark is more than an order of magnitude heavier ($m_s \sim$ 150 MeV), but still sometimes considered "light" on the typical GeV scales of strong interaction physics. It will nevertheless be ignored in the present discussion, together with the heavy quarks ($Q = c$, $b$ and $t$).  

Consider QCD in the limit of massless quarks, setting $m=0$ in Eq.(\ref{QCD}). In this limit, the QCD Lagrangian has a global symmetry related to the conserved right- or left-handedness (chirality) of zero mass spin $1/2$ particles. Introducing right- and left-handed quark fields,
\begin{equation}
\psi_{R,L} = \frac{1}{2} (1 \pm \gamma_5) \psi ,
\end{equation}
we observe that separate global unitary transformations
\begin{equation}
\psi_R \to \exp [i \theta^a_R \frac{\tau_a}{2}] \, \psi_R, \hspace{1,5cm} \psi_L
\to \exp [i \theta^a_L \frac{\tau_a}{2}] \, \psi_L ,
\end{equation}
with $\tau_a (a= 1,2,3)$ the generators of (isospin) $SU(2)$, leave ${\cal L}_{QCD}$ invariant in the limit $m \to 0$. This is the chiral $SU(2)_R \times SU(2)_L$ symmetry of QCD. It implies six conserved Noether currents,
$J^{\mu}_{R, a} = \bar{\psi}_R \gamma^{\mu} \frac{\tau_a}{2} \psi_R$ and
$J^{\mu}_{L, a} = \bar{\psi}_L \gamma^{\mu} \frac{\tau_a}{2} \psi_L$, with
$\partial_{\mu} J^{\mu}_R = \partial_{\mu} J^{\mu}_L = 0 $. It is common to
introduce the vector and axial currents,
\begin{equation}
V^{\mu}_a = J^{\mu}_{R,a} + J^{\mu}_{L,a} = \bar{\psi} \gamma^{\mu }\frac{\tau_a}{2} \psi  \, , \hspace{7mm}
A^{\mu}_a (x) = J_{R,a} - J_{L,a} = \bar{\psi} \gamma^{\mu} \gamma_5 \frac{\tau_a}{2} \psi\,  .
\end{equation}
Their corresponding charges,
\begin{equation}
Q^V_a = \int d^3 x 
~\psi^{\dagger} (x) \frac{\tau_a}{2} \psi (x)\, , \hspace{1cm}
Q^A_a = \int d^3 x 
~\psi^{\dagger} (x) \gamma_5 \frac{\tau_a}{2} \psi (x) \, ,
\label{charge}
\end{equation}
are, likewise, generators of $SU(2) \times SU(2)$.

\subsection{Spontaneous chiral symmetry breaking}
\index{spontaneously broken chiral symmetry}
\index{spontaneous symmetry breaking}
There is evidence from hadron spectroscopy that the chiral $SU(2) \times SU(2)$ symmetry 
of the QCD Lagrangian (\ref{QCD}) with $m = 0$ is spontaneously broken: for dynamical reasons of non-perturbative origin, the ground state (vacuum) of QCD has lost part of the symmetry of the
Lagrangian. It is symmetric only under the subgroup $SU(2)_V$ generated by the vector charges $Q^V$. This is the well-known isospin symmetry seen in spectroscopy and dynamics.

If the ground state of QCD were symmetric under chiral $SU(2) \times SU(2)$, both vector and axial charge operators would annihilate the vacuum: $Q^V_a |0 \rangle = Q^A_a |0 \rangle = 0$. This is the 
\tindex{Wigner-Weyl realisation} of chiral symmetry with a "trivial" vacuum. It would imply the systematic appearance of parity doublets in the hadron spectrum. For example, correlation functions of vector and axial vector currents should be identical, i.~e. $\langle 0 | V^{\mu} V^{\nu} | 0\rangle = \langle 0 | A^{\mu} A^{\nu} | 0 \rangle$. Consequently, the spectra of vector $(J^{\pi} = 1^-)$ and axial vector $(J^{\pi} = 1^+)$ mesonic excitations should also be identical. This degeneracy is not seen in nature: the $\rho$ meson mass ($m_{\rho} \simeq 0.77$ GeV) is well separated from that of the $a_1$ meson ($m_{a_1} \simeq 1.23$ GeV). Likewise, the light pseudoscalar $(J^{\pi} = 0^-)$ mesons have masses much lower than the lightest scalar $(J^{\pi} = 0^+)$ mesons.

One must conclude $Q^A_a |0 \rangle \neq 0$, that is, chiral symmetry is
spontaneously broken down to isospin: $SU(2)_R \times SU(2)_L \to
SU(2)_V$. This is the \tindex{Nambu-Goldstone realisation} of chiral symmetry. \tindex{Goldstone's theorem} says
that a spontaneously broken global symmetry implies the existence of a massless boson (the
Goldstone boson). Let us give a sketch of this theorem. If $Q^A_a | 0 \rangle \neq 0$, there must be a physical state generated by the axial charge, $|\phi_a \rangle = Q^A_a | 0 \rangle$, which is
energetically degenerate with the vacuum. Let $H_0$ be the QCD Hamiltonian
(with massless quarks) which commutes with the axial charge. Setting the ground
state energy equal to zero for convenience, we have 
$H_0 |\phi_a \rangle = Q^A_a
H_0 | 0 \rangle = 0$. Evidently $| \phi_a \rangle $ represents three massless
pseudoscalar bosons (for $N_f = 2$). They are identified with the pions.

\subsection{The chiral condensate}
\index{chiral condensate}
Spontaneous chiral symmetry breaking goes together with a qualitative rearrangement of the vacuum, an entirely non-perturbative phenomenon. The ground state is now populated by scalar quark-antiquark pairs. The corresponding ground state expectation value $\langle 0 | \bar{\psi} \psi | 0 \rangle$ is called the chiral (or quark) condensate. We frequently use the notation
\begin{equation}
\langle \bar{\psi} \psi \rangle = \langle \bar{u} u \rangle + \langle \bar{d} d
\rangle ~~.
\end{equation}
The precise definition of the chiral condensate is:
\begin{equation}
\langle \bar{\psi} \psi \rangle = -i Tr \lim_{y \to x^+} S_F (x,y)
\end{equation}
with the full quark propagator, $S_F (x,y) = -i \langle 0 | {\cal T} \psi (x)
\bar{\psi} (y) | 0 \rangle$ where ${\cal T}$ denotes the time-ordered
product. We recall Wick's theorem which states that ${\cal T} \psi (x)
\bar{\psi} (y)$ reduces to the normal product $ :\psi (x) \bar{\psi} (y):$ plus
the contraction of the two field operators. When considering the perturbative
quark propagator, $S^{(0)}_F (x,y)$, the time-ordered product is taken with
respect to a trivial vacuum for which the expectation value of $ : \bar{\psi}
\psi :$ vanishes. Long-range, non-perturbative physics is then at the origin of
a non-vanishing $\langle : \bar{\psi} \psi  : \rangle$.

(In order to establish the connection between spontaneous chiral symmetry 
breaking and the non-vanishing chiral condensate in a more formal way, 
introduce the pseudoscalar operator
$P_a (x) = \bar{\psi} (x) \gamma_5 \tau_a \psi (x)$ and derive the (equal-time) commutator relation
$[ Q^A_a, P_b] = - \delta_{ab} \bar{\psi} \psi$
which involves the axial charge $Q^A_a$ of eq.(\ref{charge}). Taking the ground state expectation value, we see that $Q^A_a | 0 \rangle \neq 0$ is indeed consistent with $\langle \bar{\psi} \psi \rangle \neq 0$.)

Let $| \pi_a (p) \rangle $ be the state vectors of the Goldstone bosons associated
with the spontaneous breakdown of chiral symmetry. Their four-momenta are
denoted $p^{\mu} = (E_p, \vec{p}\,)$, and we choose the standard normalization 
\begin{equation}\langle \pi_a (p) | \pi_b (p') \rangle = 2 E_p \delta_{ab} 
(2 \pi)^3 \delta^3(\vec{p} - \vec{p} \, ')  .
\end{equation}
Goldstone's theorem also implies non-vanishing matrix elements of the axial current (4) which
connect $| \pi_a (p) \rangle$ with the vacuum:
\begin{equation}
\langle 0 | A^{\mu}_a (x) | \pi_b (p) \rangle = i p^{\mu} f \delta_{ab} e^{-ip
\cdot x} ,
\end{equation}
where $f$ is the pion decay constant (taken here in the chiral
limit, i.~e. for vanishing quark mass). Its physical value
\begin{equation}
f_\pi = (92.4 \pm 0.3) \, MeV
\end{equation}
differs from $f$ by a small correction linear in the quark mass $m_q$.

Non-zero quark masses $m_{u,d}$ shift the mass of the Goldstone boson from zero
to the observed value of the physical pion mass, $m_{\pi}$. The connection between $m_{\pi}$ and the $u$- and $d$- quark masses is provided by 
PCAC and the Gell-Mann, Oakes, Renner (GOR) relation
\index{Gell-Mann-Oakes-Renner relation}
\cite{GOR}:
\begin{equation}
m^2_{\pi} = - \frac{1}{f^2} (m_u + m_d) \langle \bar{q} q \rangle + 
{\cal O}(m^2_{u,d}) .
\label{GOR}
\end{equation}
We have set $\langle \bar{q} q \rangle \equiv \langle \bar{u} u \rangle \simeq
\langle \bar{d} d \rangle$ making use of isospin symmetry which is valid to a
good approximation. Neglecting terms of order $m^2_{u,d}$,
identifying $f = f_{\pi} = 92.4$ MeV to this order and inserting $m_u + m_d
\simeq 12$ MeV \cite{PP,Ioffe1} (at a renormalisation scale of order 1 GeV), 
one obtains
\begin{equation}
\langle \bar{q} q \rangle \simeq - (240 \, MeV)^3 \simeq -1.8 \, fm^{-3} .
\label{cond}
\end{equation}
This condensate (or correspondingly, the pion decay constant $f_{\pi}$) is a
measure of spontaneous chiral symmetry breaking. The non-zero pion mass, on the
other hand, reflects the explicit symmetry breaking by the small quark masses,
with $m^2_{\pi} \sim m_q$. It is important to note that $m_q$ and $\langle
\bar{q} q \rangle$ are both scale dependent quantities. Only their product
$m_q \langle \bar{q} q \rangle$ is invariant under the renormalization group.

In order to appreciate the strength of the scalar condensate (\ref{cond}) from a nuclear physics point of view, note that its magnitude is more than a factor of ten larger than the baryon density in the bulk of a heavy nucleus, $\rho \simeq 0.16$ fm$^{-3}$.
 
\subsection{The nucleon mass and the gap in the hadron spectrum}

\index{QCD nucleon mass}
In the hadronic phase of QCD in which nuclei reside, confinement implies that the active ("effective") degrees of freedom are not elementary quarks and gluons but mesons and baryons. 
The observed spectrum of low-mass hadrons has a characteristic gap, $\Delta
\sim M_N \sim 1$ GeV, which separates the masses
of all baryons and almost all mesons from the ground state $| 0 \rangle$. On
the other hand, the lightest pseudoscalar mesons are positioned well within this gap, for
good reason: as Goldstone bosons of spontaneously broken chiral symmetry, pions
would start out massless. Explicit symmetry breaking by the masses $m_q$ of the
light quarks introduces perturbations on a scale small compared to $\Delta$.

The appearance of the gap $\Delta$ is thought to be closely linked to the presence of the chiral condensate $\langle \bar{\psi} \psi \rangle$ in the QCD ground
state. For example, Ioffe's formula \cite{Ioffe81}, based on QCD sum rules, connects
the nucleon mass $M_N$ directly with  $\langle \bar{\psi} \psi \rangle$  in
leading order:
\begin{equation}
M_N = - \frac{4 \pi^2}{\Lambda^2_B} \langle \bar{\psi} \psi \rangle + ...~~ ,
\label{Io}
\end{equation}
where $\Lambda_B \sim 1$ GeV is an auxiliary scale (the Borel mass) 
which separates
"short" and "long"-distance physics in the QCD sum rule analysis. While this formula is not very accurate and needs to be improved by including higher order condensates, it nevertheless demonstrates that
spontaneous chiral symmetry breaking plays an essential role in giving the nucleon its mass. 

The condensate  $\langle \bar{\psi} \psi \rangle$ is encoded in the pion decay
constant $f_{\pi}$ through the GOR relation (\ref{GOR}). In the chiral limit $(m_q \to
0)$, this $f_{\pi}$ is the only quantity which can serve to define a mass scale
("transmuted" through non-perturbative dynamics from the scale $\Lambda_{QCD}\simeq 0.2$ GeV appearing in the QCD running coupling strength). It is common to introduce $ 4 \pi f_{\pi} \sim 1$ GeV as the scale characteristic of spontaneous chiral symmetry breaking. This scale is then roughly identified with the spectral gap $\Delta$. 

At a more basic level, the nucleon mass is determined by the energy-momentum tensor $T^{\mu\nu}$ derived from the QCD Lagrangian (\ref{QCD}). The component $T^{00} = H$ defines the Hamiltonian, and $M_N = \langle N |H| N\rangle$. A detailed analysis \cite{Ji} shows that roughly $2/3$ of $M_N$ can be thought of as having its origin in the gluon field energy; the remainder results from the combination of quark kinetic and potential energies. Alternatively, one can express the nucleon mass as the expectation value $M_N = \langle N|T_\mu^{\mu} |N \rangle$ of the non-vanishing trace
\begin{eqnarray}
T_\mu^{\mu}&=&\frac{\beta(g)}{2g}G_{\mu \nu}G^{\mu \nu} + m_u \,\bar{u}u+m_d\,\bar{d}d+... ~~  ,
\end{eqnarray}
of the energy-momentum tensor (the so-called trace anomaly - see e.g. ref.\cite{DGH}). Here $G^{\mu \nu}$ is the gluonic field strength tensor and $\beta(g)$ is the beta function of QCD ($\beta = - \beta_0g^3/(4\pi)^2$ in leading order, with $\beta_0 = 11- 2N_f/3$). The quark mass terms\footnote{We have omitted here the anomalous dimension of the mass operator, as in \cite{Ji}.}  $m_q \bar{q}q$ with $q=u,\,d \dots$ include light as well as heavier quarks.  
The physical nucleon mass $M_N$ is expressed as
\begin{eqnarray}
M_N&=& M_0+\sigma_N
\end{eqnarray}
in terms of its evidently non-vanishing value in the chiral limit,
\begin{eqnarray}
M_0&=&\langle N |\frac{\beta}{2g} G_{\mu \nu}G^{\mu \nu} + \, ... \,| N \rangle 
\label{mass}
\end{eqnarray}
(where the dots refer to possible contributions from heavier quarks, other than $u$ and $d$), and the sigma term
\index{nucleon sigma term}
\begin{eqnarray}
\sigma_N&=&\! \sum_{q=u,d}  m_q \frac{{\rm{d}} M_N}{{\rm{d}}m_q} = \langle N |m_u\, \bar{u}u+m_d\,   \bar{d}d| N \rangle     .
\label{sigma}
\end{eqnarray}
This sigma term represents the contribution from explicit chiral symmetry breaking to the nucleon mass, through the small but non-vanishing $u$- and $d$-quark masses.  

Relations such as Eq.(\ref{Io}), when combined with Eq.(\ref{mass}), give
important hints. Non-perturbative gluon dynamics confines quarks and generates hadron masses, producing a gap in the spectrum. In the sector of the (almost) massless $u$- and $d$- quarks, this gap reflects the spontaneously broken chiral symmetry. Systems characterized by an energy gap usually exhibit qualitative changes when exposed to variations of thermodynamic conditions. A
typical example is the temperature dependence of the gap in a
superconductor. For the physics in the hadronic phase of QCD, the following key
issues need therefore to be addressed: how does the quark condensate 
$\langle \bar{\psi}\psi \rangle$ 
change with temperature and/or baryon density? Are there
systematic changes of hadronic spectral functions in a dense and hot medium,
which would indicate changes of the QCD vacuum structure? How does the nucleon mass itself
vary under such conditions? And what is the impact of these considerations on nuclear many-body
systems? The sigma term (\ref{sigma}) will play an important role in this discussion as it controls the
leading density dependence of the chiral condensate (see Section 2.7).

\subsection{Chiral effective field theory}

\index{chiral effective field theory}
The mass scale given by the gap $\Delta \sim 4 \pi f_{\pi}$ offers a
natural separation between "light" and "heavy" (or, correspondingly, "fast" and
"slow") degrees of freedom. The basic idea of an effective field theory is to
introduce the active light particles as collective degrees of freedom,  while the
heavy particles are frozen and treated as (almost) static sources. The dynamics
is described by an effective Lagrangian which incorporates all relevant
symmetries of the underlying fundamental theory. We now list the necessary
steps, first for the pure meson sector (baryon number $B = 0)$ and later for
the $B = 1$ sector. We work mostly with two quark flavours ($N_f = 2$) in the following.

a) The elementary quarks and gluons of QCD are replaced by Goldstone
bosons. They are represented by a $2 \times 2$ matrix field $U (x) \in
SU(2)$ which collects the three isospin components $\pi_a (x)$ of 
the Goldstone pion. A
convenient choice of coordinates is
\begin{equation}
U (x) = \exp[i \tau_a \phi_a(x)]~~ ,
\end{equation}
with $\phi_a = \pi_a/f$ where the pion decay constant $f$ in the chiral limit
provides a suitable normalisation. (Other choices, such
as $U = \sqrt{1- \pi^2_a / f^2} + i \tau_a \pi_a /f$, are also common.
Results obtained from the effective theory must be independent of the
coordinates used.)

b) Goldstone bosons interact weakly at low energy. In fact, if $| \pi
\rangle = Q^A | 0 \rangle$ is a massless state with $H | \pi \rangle = 0$, then
a state $| \pi^n \rangle = (Q^A)^n | 0 \rangle$ with $n$ Goldstone bosons is
also massless since the axial charges $Q^A$ all commute with the full
Hamiltonian $H$. Interactions between Goldstone bosons must therefore vanish at
zero momentum and in the chiral limit.

c) The QCD Lagrangian (\ref{QCD}) is replaced by an effective Lagrangian which
involves the field $U (x)$ and its derivatives:
\begin{equation}
{\cal L}_{QCD} \to {\cal L}_{eff} (U, \partial U, \partial^2 U, ...) .
\label{Lagr}
\end{equation}
Goldstone bosons do not interact unless they carry non-zero four-momentum, so the low-energy
expansion of (\ref{Lagr}) is an ordering in powers of $\partial_{\mu} U$. Lorentz
invariance permits only even numbers of derivatives. We write
\begin{equation}
{\cal L}_{eff} = {\cal L}_\pi^{(2)} + {\cal L}_\pi^{(4)} + ...
\end{equation}
and omit an irrelevant constant term. The leading term, called non-linear sigma
model, involves two derivatives:
\begin{equation}
{\cal L}_\pi^{(2)} = \frac{f^2}{4} Tr [ \partial_{\mu} U^{\dagger} \partial^{\mu} U
] .
\end{equation}
At fourth order, the terms permitted by symmetries are (apart from an extra
contribution from the QCD anomaly, not included here):
\begin{equation}
{\cal L}_\pi^{(4)} = \frac{l_1}{4} ( Tr [\partial_{\mu} U^{\dagger}
\partial^{\mu} U])^2 +
\frac{l_2}{4} Tr [\partial_{\mu} U^{\dagger} \partial_{\nu} U] Tr
[\partial^{\mu} U^{\dagger} \partial^{\nu} U] ,
\end{equation}
and so forth. The constants $l_1, l_2$ (following canonical notations of
ref.~\cite{GL}) must be determined by experiment. To the extent that the
effective Lagrangian includes all terms dictated by the symmetries, the
effective theory is equivalent to QCD \cite{WL}.

d) The symmetry breaking mass term is small, so that it can be handled
perturbatively, together with the power series in momentum. The leading
contribution introduces a term linear in the quark mass matrix $m$:
\begin{equation}
{\cal L}_\pi^{(2)} = \frac{f^2}{4} Tr [\partial_{\mu} U^{\dagger} \partial^{\mu} U]
+ \frac{f^2}{2} B_0 \, Tr [m (U + U^{\dagger})] .
\end{equation}
The fourth order term ${\cal
L}_\pi^{(4)}$ also receives symmetry breaking contributions with additional constants
$l_i$.

When expanding ${\cal L}^{(2)}$ to terms quadratic in the pion field, one finds
\begin{equation}
{\cal L}_\pi^{(2)} = (m_u + m_d) f^2 B_0 + \frac{1}{2} \partial_{\mu} \pi_a
\partial^{\mu} \pi_a - \frac{1}{2} (m_u + m_d) B_0 \pi^2_a + 0 (\pi^4) .
\label{L2}
\end{equation}
At this point we can immediately verify the GOR relation (\ref{GOR})
at the level of the effective theory. 
The first (constant) term in (\ref{L2}) corresponds to the
shift of the vacuum energy density by the non-zero quark masses. Identifying
this with the vacuum expectation value of the corresponding piece in the QCD
Lagrangian (\ref{QCD}), we find $- (m_u \langle \bar{u} u \rangle + m_d \langle \bar{d}
d \rangle) = (m_u + m_d) f^2 B_0$ and therefore $\langle \bar{u} u \rangle =
\langle \bar{d} d \rangle = - f^2 B_0$ in the chiral limit, $m_{u,d} \to 0$. 
The pion mass term in (\ref{L2}) is evidently identified as $m^2_{\pi} = (m_u + m_d)
B_0$. Inserting $B_0$, we have the GOR relation.

e) Given the effective Lagrangian, the framework for systematic perturbative
calculations of the S-matrix involving Goldstone bosons, named Chiral
Perturbation Theory (ChPT), is then defined by the following rules: 
\index{chiral perturbation theory (ChPT)}

Collect all Feynman diagrams generated by ${\cal L}_{eff}$. Classify
all terms according to powers of a variable $Q$
which stands generically for three-momentum or energy 
of the Goldstone bosons, or for the pion mass $m_{\pi}$. The small
expansion parameter is $Q/4\pi f_{\pi}$. Loops are subject
to dimensional regularization and renormalization. 

\subsection{Effective Lagrangian including baryons}

The prominent role played by the pion as a Goldstone boson of spontaneously
broken chiral symmetry has its strong impact on the low-energy structure
and dynamics of nucleons as well. Decades of research in nuclear physics
have established the pion as the mediator of the long-range force between
nucleons. When probing the individual nucleon itself with long-wavelength
electroweak fields, a substantial part of the response comes from the pion
cloud, the ``soft'' surface of the nucleon. 

The calculational framework for this, 
\tindex{baryon chiral perturbation theory}, 
has been applied quite successfully to a variety of low-energy processes
(such as threshold pion photo- and electroproduction and Compton scattering on 
the nucleon) for which increasingly accurate experimental data have become
available in recent years. An introductory survey is given in chapter 7
of ref.~\cite{TW}. A detailed review can be found in ref.~\cite{BKM}.
 
Let us consider the sector with baryon number $B = 1$ and concentrate
on the physics of the pion-nucleon system, restricting ourselves to the
case of $N_f = 2$ flavours. 

The nucleon is represented by a Dirac spinor field $\Psi_N(x) = (p,n)^T$ 
organised as an isospin-$1/2$ doublet of proton and neutron. The free field
Lagrangian
\begin{equation}
{\cal L}_0^N = \bar{\Psi}_N(i\gamma_{\mu}\partial^{\mu} - M_0)\Psi_N
\label{Nuc}
\end{equation}
includes the nucleon mass in the chiral limit, $M_0$. Note that nucleons, 
unlike pions, supposedly have at least part of their large mass 
connected with the strong scalar mean field provided by the quark condensate 
$\langle\bar{\psi}\psi\rangle$. Such a relationship is explicit, for example,
in the Ioffe formula (\ref{Io}).

We can now construct the low-energy effective Lagrangian for pions 
interacting with a nucleon. The previous pure meson Lagrangian
is replaced by ${\cal L}_{eff} = {\cal L}_{\pi}^{(2)} + 
{\cal L}_{\pi}^{(4)} + ... + {\cal L}_{eff}^N $ which also
includes the nucleon field. The additional term involving the nucleon is 
expanded again in powers of derivatives (external
momenta) and quark masses: 
\begin{equation}
{\cal L}_{eff}^N = {\cal L}_{\pi N}^{(1)} + {\cal L}_{\pi N}^{(2)} ~...
\end{equation}
Let us discuss the leading term, ${\cal L}_{\pi N}^{(1)}$. The modifications
as compared to the free nucleon Lagrangian (\ref{Nuc}) are twofold. First, there is
a replacement of the $\partial^{\mu}$ term by a chiral covariant derivative
which introduces vector current couplings between the pions and the nucleon.
Secondly, there is an axial vector coupling. This structure of the $\pi N$
effective Lagrangian is again dictated by chiral symmetry. We have
\begin{equation}
{\cal L}_{\pi N}^{(1)} =  \bar{\Psi}_N[i\gamma_{\mu}(\partial^{\mu} 
- iv^{\mu}) + \gamma_{\mu}\gamma_5 a^{\mu} - M_0]\Psi_N~~,
\label{LPN}
\end{equation}
with vector and axial vector quantities involving the Goldstone boson (pion)
fields in the form $\xi = \sqrt{U}$:
\begin{eqnarray}
v^{\mu} & = & \frac{i}{2}(\xi^{\dagger}\partial^{\mu}\xi +  
\xi\partial^{\mu}\xi^{\dagger}) = -\frac{1}{4f^2}\varepsilon_{abc}\tau_a~\pi_b
~\partial^{\mu}\pi_c + ...~~, \\
a^{\mu} & = & \frac{i}{2}(\xi^{\dagger}\partial^{\mu}\xi -  
\xi\partial^{\mu}\xi^{\dagger}) = -\frac{1}{2f^2}\tau_a~
\partial^{\mu}\pi_a + ...~~,
\end{eqnarray}
where the last steps result when expanding $v^{\mu}$ and $a^{\mu}$ to 
leading order in the pion fields. 

So far, the only parameters that enter are the nucleon mass, $M_0$,
and the pion decay constant, $f$, both taken in the chiral limit and ultimately
connected with a single scale characteristic of non-perturbative QCD and 
spontaneous chiral symmetry breaking.

When adding electroweak interactions to this scheme, one observes an additional
feature which has its microscopic origin in the substructure of the nucleon,
not resolved at the level of the low-energy effective theory. The analysis of
neutron beta decay $(n \rightarrow p e \bar{\nu})$ reveals that the 
$\gamma_{\mu}\gamma_5$ term in (\ref{LPN}) is to be multiplied by the axial vector
coupling constant $g_A$, with the empirical value
\begin{equation}
g_A = 1.267 \pm 0.003~~.
\end{equation}

At next-to-leading order $({\cal L}_{\pi N}^{(2)})$, the symmetry breaking 
quark mass term enters. It has the effect of shifting the nucleon mass from
its value in the chiral limit to the physical one, introducing the sigma term $\sigma_N$
of Eq.(\ref{sigma}). The empirical sigma term,
\index{nucleon sigma term}
\begin{equation}
\sigma_N = (45 \pm 8) MeV\,\,,
\label{sigval}
\end{equation}
has been deduced by a sophisticated extrapolation of low-energy pion-nucleon data
using dispersion relation techniques \cite{GLS}. There is still an ongoing debate, related
to different ways of handling the $\pi N$ phase shift analysis, whether
$\sigma_N$ could in fact be larger than that given in (\ref{sigval}). On the other hand, a recent 
extrapolation of two-flavour lattice QCD results using methods of chiral effective field theory \cite{PHW}
yields\footnote{note: statistical errors only; systematic lattice uncertainties from finite volume effects etc. still require further detailed studies.} $\sigma_N = (47 \pm 3)$ MeV, perfectly consistent with Eq.(\ref{sigval}).

Up to this point, the $\pi N$ effective Lagrangian, expanded to second order
in the pion field, has the form
\begin{eqnarray}
{\cal L}_{eff}^{N} & = & \bar{\Psi}_N(i\gamma_{\mu}\partial^{\mu} 
- M_N)\Psi_N - \frac{g_A}{2f_{\pi}} \bar{\Psi}_N\gamma_{\mu}\gamma_5 
\mbox{\boldmath $\tau$}\Psi_N\cdot\partial^{\mu}\mbox{\boldmath $\pi$}  \\
                   &   & -\frac{1}{4f_{\pi}^2}
 \bar{\Psi}_N\gamma_{\mu} 
\mbox{\boldmath $\tau$}\Psi_N\cdot\mbox{\boldmath $\pi$}\times
\partial^{\mu}\mbox{\boldmath $\pi$}
+ \frac{\sigma_N}{f_{\pi}^2} \bar{\Psi}_N\Psi_N\mbox{\boldmath $\pi$}^2 
+ ...~~,\nonumber
\end{eqnarray}
where we have not shown a series of additional terms of order 
$(\partial^{\mu} {\bm{\pi}})^2$ included in the complete 
${\cal L}_{\pi N}^{(2)}$.
These terms come with further constants that need to be fitted to experimental
data. In later applications using in-medium chiral perturbation theory, the
calculations will be based on the leading terms, linear in the derivative
$\partial^{\mu} {\bm{\pi}}$ of the pion field, while the term involving 
$\sigma_N$
will be manifest in the density dependence of the chiral condensate, as we
shall see now.

\subsection{Chiral thermodynamics}

\index{chiral thermodynamics}
Before turning our attention to the nuclear many-body system, it is of some relevance
to examine the thermodynamics based on the chiral effective Lagrangian ${\cal L}_{eff} =
\sum_i ({\cal L}_\pi^{(i)} + {\cal L}_{\pi N}^{(i)})$. In particular, we are interested in the question how
the quark condensate $\langle \bar{q}q \rangle$ behaves with changing thermodynamic conditions.

\subsubsection{A. Thermodynamics of the chiral condensate}

\index{chiral condensate}
Consider as a starting point the partition function
\begin{equation}
{\cal Z} = Tr \exp\left[-\frac{1}{T}\int_V d^3x~({\cal H}-\mu\rho)\right]~~,
\end{equation}
where ${\cal H}$ is the Hamiltonian density, $\mu$ denotes the chemical 
potential, and $\rho$ the baryon density. The partition function at $\mu = 0$, expressed in terms of the QCD Hamiltonian
$H$, is
\begin{equation}
{\cal Z} = Tr \exp (-H/T) = \sum_n \langle n | e^{-E_n /T} | n \rangle
\label{partition}
\end{equation}
with $(H - E_n) | n \rangle = 0$. Confinement implies that the eigenstates $| n \rangle$ of $H$ are (colour-singlet) hadrons at low temperatures, below the critical $T_c \sim 0.2$ GeV  for the deconfinement transition. The physics is then determined by the states of lowest mass in the spectrum $\{ E_n\}$, namely the pions. At non-zero baryon chemical potential $\mu$, nucleons enter in addition as the lowest-mass baryons. The Hamiltonian density  ${\cal H}$ of QCD is expressed in terms of the relevant low-energy degrees of freedom in the hadronic phase, derived from the chiral effective Lagrangian ${\cal L}_{eff}$. The quark mass term,
\begin{equation}
\delta{\cal H} = \bar{\psi}m\psi = m_u\,\bar{u}u + m_d\,\bar{d}d\,\,+\,\,...~~,
\end{equation}
can be separated so that ${\cal H} = {\cal H}_0 + \delta{\cal H}$, with ${\cal H}_0$
representing the massless limit.

Now assume a homogeneous medium and consider the pressure (or the free
energy density)
\begin{equation}
P(T,V,\mu) = -{\cal F}(T,V,\mu) = \frac{T}{V}ln{\cal Z}~~.
\end{equation}
The derivative of $P$ with respect to a quark mass $m_q$ of given
flavour $q = u,d$ obviously produces the in-medium quark
condensate, the thermal expectation value $\langle \bar{q}q\rangle_T$
as a function of temperature and chemical potentials. Subtracting
vacuum quantities one finds
\begin{equation}
\langle \bar{q}q\rangle_{T,\rho} = \langle \bar{q}q\rangle_0
- \frac{dP(T,V,\mu)}{dm_q}~~,
\label{con}
\end{equation}
where $\langle \bar{q}q\rangle_0$ refers to the vacuum condensate
taken at $T = 0$ and $\mu = 0$. The $\mu$-dependence of the condensate
is converted into a density dependence via the relation $\rho = \partial
P/\partial\mu$ at fixed $T$.
Using the Gell-Mann, Oakes, Renner relation (\ref{GOR}), one can rewrite 
Eq.(\ref{con}) as
\begin{equation}
\frac{\langle \bar{q}q\rangle_{T,\rho}}{\langle \bar{q}q\rangle_0} =
1 + \frac{1}{f_{\pi}^2}\frac{dP(T,\mu)}{dm_{\pi}^2}~~.
\label{con2}
\end{equation}
The task is therefore to investigate how the equation of state changes,
at given temperature and baryon chemical potential, when varying the 
quark mass (or, equivalently, the squared pion mass).

For a system of nucleons interacting with pions, the total 
derivative of the pressure with respect to $m_{\pi}^2$ reduces to
\begin{equation}
\frac{dP}{dm_{\pi}^2} = \frac{\partial P}{\partial m_{\pi}^2}
+ \frac{\partial M_N}{\partial m_{\pi}^2}\frac{\partial P}{\partial M_N}
= \frac{\partial P}{\partial m_{\pi}^2} - \frac{\sigma_N}{m_{\pi}^2}
\rho_s(T,\mu)~~,
\label{DP}
\end{equation}
with the scalar density $\rho_s = -\partial P/\partial M_N$ and the 
sigma term $\sigma_N = m_{\pi}^2 \partial M_N/\partial m_{\pi}^2$ translated from Eq.(\ref{sigma}) using
the GOR relation.

Next, consider the limit of low baryon density $\rho$ at zero temperature,
$T = 0$. In this case nucleons are the only relevant degrees of freedom.
At sufficiently low density, i.e. when the average distance between two 
nucleons is much larger than the pion Compton wavelength, the nucleon mass remains
at its vacuum value $M_N$, and one can neglect $NN$ interactions. The pressure
is that of a Fermi gas of nucleons\footnote{Actually, this statement needs to be modified at extremely low densities where the system turns into a gas of clusters (deuterons etc.)}, subject only to the Pauli principle.
Returning to Eqs.(\ref{con2},\ref{DP}), we have
\begin{equation}
\frac{\langle \bar{q}q\rangle_\rho}{\langle \bar{q}q\rangle_0} =
1 - \frac{\sigma_N}{m_\pi^2 f_\pi^2}\rho_s~~,
\label{mediumcond}
\end{equation}
with the scalar density
\begin{equation}
\rho_s = -\frac{\partial P}{\partial M_N} =
4\int_{|\vec{p}|\leq k_f} \frac{d^3p}{(2\pi)^3}
\frac{M_N}{\sqrt{\vec{p}\,^2 + M_N^2}}~~,
\end{equation}
for a system with equal number of protons and neutrons, i.e. with degeneracy
factor $d=4$ from spin and isospin. The Fermi momentum $k_f$ is related to
the baryon density by $\rho = 2k_f^3/(3\pi^2)$.
At low baryon densities with $k_f^2 \ll M_N^2$, the difference between $\rho_s$
and $\rho$ can be neglected, so that
\begin{equation}
\frac{\langle \bar{q}q\rangle_\rho}{\langle \bar{q}q\rangle_0} \approx
1 - \frac{\sigma_N}{m_\pi^2 f_\pi^2}\rho~~.
\label{lin}
\end{equation}
The consequences implied by this relation are quite remarkable. 
Using the empirical value $\sigma_N\simeq 50$ MeV of the nucleon 
sigma term one observes that the chiral condensate at normal nuclear
matter density, $\rho = \rho_0 = 0.16~fm^{-3}$, is expected to 
decrease in magnitude to less than $2/3$ of its vacuum value, a significant 
effect that should have observable consequences already in the bulk parts
of ordinary nuclei.

In the limit of low temperature and at vanishing baryon 
density $(\rho = 0)$, pions are the only thermally active degrees of
freedom in the partition function (\ref{partition}). This case is well described
by chiral perturbation theory. The combined low-temperature and low-density
behaviour is summarised as
\begin{equation}
\frac{\langle \bar{q}q\rangle_\rho}{\langle \bar{q}q\rangle_0} \approx
1 - \frac{T^2}{8f_{\pi}^2} - 0.37\left(\frac{\sigma_N}{50 MeV}\right)
\frac{\rho}{\rho_0} + ... ~~,
\end{equation}
showing a rather weak leading dependence of the chiral condensate 
on temperature, whereas its density dependence is far more pronounced.

\subsubsection{B. A schematic model}

Consider now a schematic model for the hadronic phase of QCD, starting
from an effective Lagrangian ${\cal L}_{eff} = {\cal L}_\pi + {\cal L}_{\pi N} + {\cal L}_{NN}$ in which
${\cal L}_\pi$ is given by Eq.(\ref{L2}), ${\cal L}_{\pi N}$ is the pion-nucleon Lagrangian of Eq.(\ref{LPN}) and additional short distance interactions between nucleons are represented by $NN$ contact terms,
\begin{equation}
{\cal L}_{NN} = -\frac{G_S}{2}(\bar{\Psi}_N\Psi_N)^2 - 
\frac{G_V}{2}(\bar{\Psi}_N\gamma_{\mu} \Psi_N)^2 + ...~~.
\end{equation}  
Their coupling strength parameters $G_S < 0$ and $G_V > 0$ are fixed to reproduce ground state
properties of normal nuclear matter. What we have in mind here 
is a variant of relativistic mean field theory combined with 
``soft'' pion fluctuations treated in the framework of chiral perturbation theory
and to be discussed in more detail later.
\begin{figure}[t]
\begin{center}
\parbox{10cm}{
\includegraphics[width=10cm,clip=true]{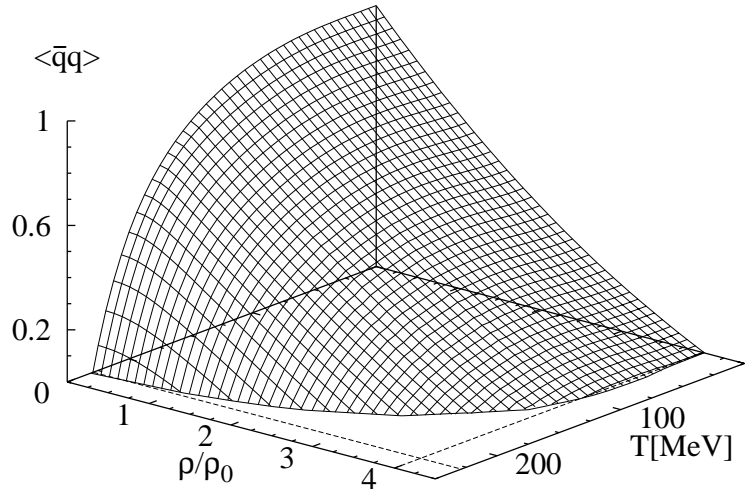}}
\end{center}
\caption{Chiral condensate (in units of its vacuum value) as a function
of temperature and baryon density ($\rho_0 = 0.16$ fm$^{-3}$ is the density
of normal nuclear matter).}
\label{fig:qqT}
\end{figure}

Using two-loop thermal field theory in order to perform a self-consistent
calculation of the pressure $P(T,\mu)$ in this model, one can deduce  
the chiral condensate as a function of temperature and baryon density
following Eq.(\ref{con2}). This calculation generates temperature dependent
mean fields for the nucleons at the same time as it treats thermal pion
fluctuations with leading $\pi\pi$ interactions. The pressure equation
takes the form
\begin{equation}
P(T,\mu) = P_N(T,\mu^*,M_N^*) + P_\pi(T,\mu^*,M_N^*)
+\frac{G_S}{2}\rho_s^2 + 
\frac{G_V}{2}\rho^2~~~,
\end{equation}
with nucleon and pion contributions $P_{N,\pi}$ respectively depending on
the effective nucleon mass $M_N^* = M_N + G_S\,\rho_S$
and the shifted baryon chemical potential $\mu^* = \mu - G_V~\rho$.
The baryon and scalar densities are determined as
\begin{equation}
\rho = \frac{\partial P}{\partial\mu} =
\frac{\partial(P_N + P_\pi)}{\partial\mu^*}, ~~~~
\rho_s = \frac{\partial P}{\partial M_N} =
\frac{\partial(P_N + P_\pi)}{\partial M_N^*}~~~.  
\end{equation}
Fixing $G_{S,V}$ to the energy per particle and the equilibrium 
density of cold nuclear matter, one can map out the nuclear equation
of state first at low temperature and density, reproducing known physics
in this domain which is a prerequisite for extrapolating into more extreme
regions.

We now return to the evaluation of the chiral condensate. The dependence
of $P(T,\mu)$ on the pion mass is explicit in the thermal pion Green
function and implicit through the nucleon mass. The result \cite{Schw} 
for $\langle \bar{q}q\rangle_{T,\rho}$ is shown in Fig.\ref{fig:qqT}. 
At low baryon density the linear behaviour of Eq.(\ref{lin}) is recovered. 
Pionic fluctuations, calculated up to three-loop order with inclusion
of two-pion exchange effects, help maintain this approximately
linear dependence not only up to $\rho\simeq\rho_0$, but even slightly beyond.
Up to this point we can conclude that the magnitude of the quark
condensate at normal nuclear matter density is expected to be reduced by 
about one third from its vacuum value, whereas the temperature dependence
is far less pronounced, at least up to $T\le 100$ MeV.
 
\section{Chiral dynamics and the nuclear many body problem}

Nuclei are aggregates of nucleons (and mesons) which are in turn clusters
of quarks and gluons. Nuclei are thus central cornerstones in the 
QCD phase diagram. The decription of nuclear many-body systems must 
ultimately be linked to
and constrained by QCD. Recent developments using
effective field theory methods are aiming to establish such connections. 
We focus here on a framework that relates strongly to 
principles of spontaneous chiral symmetry breaking in QCD and to the topics 
presented in the previous sections. 

We propose a two-step approach. The first step emphasises the
role of pions in nuclear many-body dynamics, given the rules
dictated by chiral symmetry. Our strategy is based on in-medium chiral 
perturbation theory where the nuclear Fermi momentum enters as an additional 
scale in the systematic low-energy expansion of chiral effective field theory.
As will be demonstrated, this approach is already capable of producing binding
and saturation of nuclear matter. At the same time it gives a remarkably
good value for the asymmetry energy with no additional parameter. At that
level, however, important features such as the nuclear spin-orbit force cannot yet be understood.
The second step is to provide strong scalar and vector mean fields, about
equal in magnitude but opposite in sign, which are known phenomenologically
to account for the large spin-orbit splitting in finite nuclei.
These strong mean fields will be viewed as arising from
the changes of the condensate structure of the QCD vacuum in the presence
of the nuclear medium. QCD sum rules at finite baryon density are used for guidance to
set constraints on the in-medium modifications of the relevant condensates.

\subsection{Nuclear matter, part I}

The present status of the nuclear matter problem can briefly be summarised
as follows: a quantitatively
successful description is achieved, using advanced many-body techniques
\cite{nonrel}, in a non-relativistic framework when invoking an adjustable
three-body force. Complementary to such an ab-initio approach,
relativistic mean field phenomenology, including
non-linear terms with adjustable parameters, have also been widely used for the
calculation of nuclear matter properties and 
finite nuclei \cite{SW.97}. At a more
basic level, the Dirac-Brueckner method \cite{rolf} solves a relativistically
improved Bethe-Goldstone equation with one-boson exchange $NN$-interactions.
The resulting Brueckner G-matrix is then used as input in Hartree-Fock calculations.

In recent years a novel approach to the $NN$ interaction based on effective
field theory (in particular, chiral perturbation theory) has emerged
\cite{ksw,epel,nnpap1,nnpap2}. Its key element is a power counting scheme which separates long-
and short-distance dynamics. Methods of effective field theory have also been
applied to systems of finite density \cite{furn,hammer}.

In the following sections the primary purpose is to point out the 
importance of explicit
pion dynamics in the nuclear many-body problem. While pion exchange processes
are well established as generators of the long and intermediate range
$NN$ interaction, their role in nuclear matter is less evident. The one-pion
exchange Hartree term vanishes identically for a spin-saturated system, 
and the leading
Fock exchange term is small. Two-pion exchange mechanisms are commonly hidden 
behind a purely phenomenological scalar ("sigma") field which is 
fitted to empirical data but has no deeper justification. This situation calls
for a more detailed understanding. We report on steps in this direction,
following ref. \cite{KFW1}. A conceptually similar strategy has been proposed in ref. \cite{LFA}.

Before passing on to explicit calculations it is useful to draw attention to 
the following fact. A simple but surprisingly good parametrization of the energy per
particle, $\bar{E}(k_f) = E(k_f)/A$, of isospin symmetric nuclear matter is given in
powers of the Fermi momentum $k_f$ as
\begin{equation}
\bar{E} (k_f) = \frac{3 k^2_f}{10 \, M_N} - \alpha \frac{k^3_f}{M_N^2} + \beta
\frac{k^4_f}{M_N^3} ,
\label{simple_eos}
\end{equation}
where the nucleon density is $\rho = 2 k^3_f / 3 \pi^2$ as usual, and $M_N$ 
is the free nucleon mass. The first term is the kinetic energy of
a Fermi gas. Adjusting the (dimensionless) parameters $\alpha$ and $\beta$ to
the equilibrium density, $\rho_0 = 0.16$ fm$^{-3}$, ($k_{f0} = 1.33$ 
fm$^{-1}$) and $\bar{E}_0 = \bar{E} (k_{f 0)} = -16$ MeV, gives $\alpha
= 5.27$ and $\beta = 12.22$. The compression modulus $K_0 = k_{f 0}^2
(\partial^2 \bar{E} (k_f) / \partial k^2_f)_{k_{f0}}$ is then predicted at $K_0
= 236$ MeV, well in line with empirically deduced values, and the density
dependence of $\bar{E}(k_f)$ using eq. (\ref{simple_eos})
is remarkably close to the one 
resulting from the realistic many-body calculations of the Urbana group
\cite{FP.81}. The expansion (\ref{simple_eos}) 
in powers of Fermi momentum (rather than
in powers of density) is the natural one in the context of in-medium
chiral effective field theory to which we now turn our attention.
 
\subsection{In-medium chiral perturbation theory}

\index{in-medium chiral perturbation theory}
The tool to investigate the implications of spontaneous and explicit chiral
symmetry breaking in QCD is chiral perturbation theory. Observables are
calculated within the framework of an effective field theory of Goldstone
bosons (pions) interacting with the lowest-mass baryons (nucleons), 
as described in sections 2.5 and 2.6. The
diagrammatic expansion of this low-energy theory in the number of loops has a
one-to-one correspondence to a systematic expansion of observables in small
external momenta and the pion (or quark) mass.

In nuclear matter, the relevant momentum scale is the Fermi momentum $k_f$. At
the empirical saturation point, $k_{f 0} \simeq 2m_{\pi}$, so the Fermi
momentum and the pion mass are of comparable magnitude at the densities of 
interest. This immediately implies that pions should be included as {\it explicit} degrees
of freedom: their propagation in matter is relevant. Pionic effects cannot be
accounted for simply by adjusting coefficients of local $NN$ contact
interactions.

Both $k_f$ and $m_{\pi}$ are small compared to the characteristic
chiral scale, $4 \pi f_{\pi} \simeq 1.2 \, GeV$. 
Consequently, the equation of state of
nuclear matter as given by chiral perturbation theory will be represented as an
expansion in powers of the Fermi momentum. The expansion coefficients are
non-trivial functions of $k_f/ m_{\pi}$, the dimensionless ratio of the two
relevant scales inherent to the problem.

The chiral effective Lagrangian (\ref{LPN}) 
generates the basic pion-nucleon coupling 
terms: the Tomozawa-Weinberg $\pi \pi NN$ contact vertex, $(1/4 f^2_{\pi}) (q^{\mu}_b -
q^{\mu}_a) \gamma_{\mu} \epsilon_{abc} \tau_c$, and the pseudovector $\pi NN$
vertex, $(g_A / 2 f_{\pi}) q^{\mu}_a \gamma_{\mu} \gamma_5 \tau_a$, where
$q_{a,b}$ denote (outgoing) pion four-momenta and $g_A$ is the axial vector
coupling constant (we choose $g_A = 1.3$ so that the Goldberger-Treiman 
relation
$g_{\pi N} = g_A M /f_{\pi}$ gives the empirical $\pi N$ coupling constant,
$g_{\pi N} = 13.2$).
The only new ingredient in performing calculations at
finite density (as compared to evaluations of scattering processes in vacuum)
is the in-medium nucleon propagator. For a relativistic nucleon with
four-momentum $p^{\mu} = (p_0, \vec{p} \,)$ it reads
\begin{equation}
(\not \!p + M) \left\{ \frac{i}{p^2 - M^2 + i \varepsilon} - 2 \pi \delta (p^2
- M^2) \theta (p_0) \theta(k_f - | \vec{p}\, |) \right\}.
\end{equation}
The second term is the medium insertion which accounts for the fact that the
ground state of the system has changed from an "empty" vacuum to a filled Fermi
sea of nucleons. Diagrams can then be organized systematically  in the number
of medium insertions, and an expansion is performed in leading inverse powers
of the nucleon mass, consistently with the $k_f$-expansion.

Our "inward-bound" strategy \cite{KFW1} is now as follows. One starts at large
distances (small $k_f$) and systematically generates the pion-induced
correlations between nucleons as they develop with decreasing distance
(increasing $k_f$). The calculations of the energy density reported here 
are performed to 3-loop order
(including terms up to order $k^5_f$) and incorporate one- and two-pion
exchange processes. The procedure involves one single momentum space cutoff
$\Lambda$ which encodes dynamics at short distances not resolved explicitly in
the effective low-energy theory. This high-momentum scale $\Lambda$ is the only free
parameter at this stage. It has to be fine-tuned. (Alternatively and equivalently, one
could use dimensional regularization, remove divergent loop integrals and 
replace them by
adjustable $NN$ contact terms which then parametrize the unresolved short-distance
physics).

\begin{figure}[t]
\begin{center}
\includegraphics[scale=0.5]{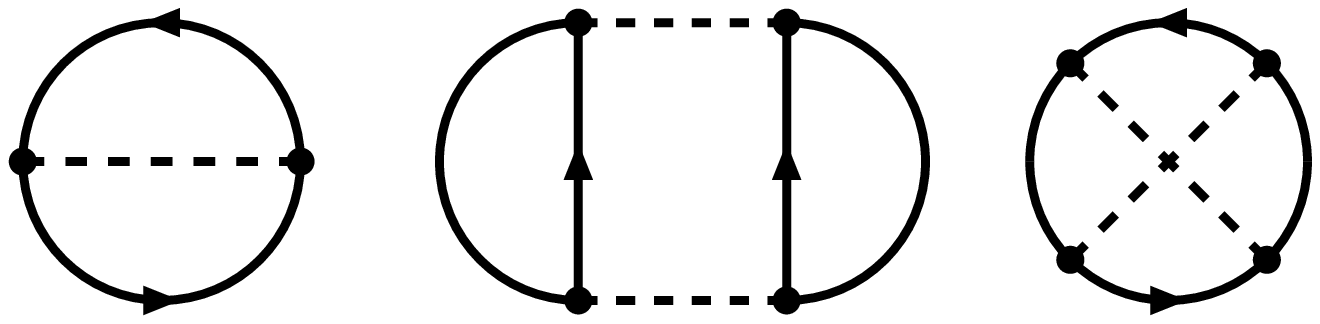} \\[1em]
\includegraphics[scale=0.5]{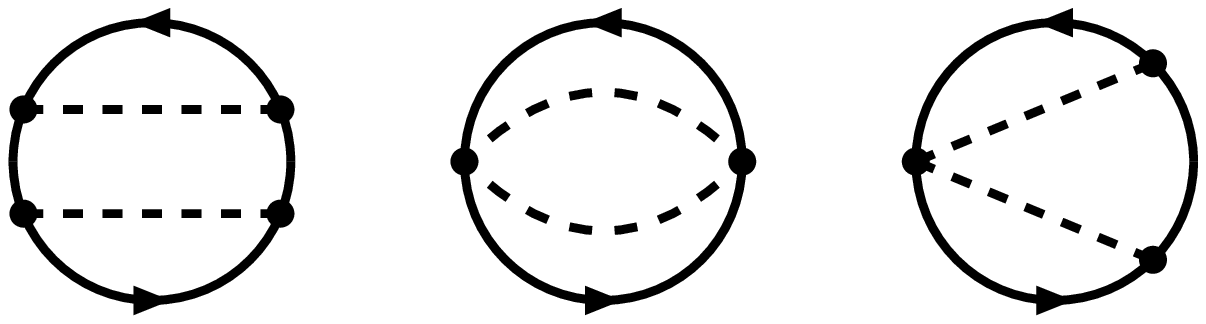}                                       
\end{center}
\caption{In-medium chiral perturbation theory: One-pion exchange Fock term
(top left), iterated one-pion exchange (top middle and right) and examples
of irreducible two-pion exchange terms (bottom). 
See Ref. \protect\cite{KFW1} for details.}
\label{VWfig1}
\end{figure}

Nuclear chiral dynamics up to three-loop order introduces the diagrams,
Fig.~\ref{VWfig1}.
They include the one-pion exchange (OPE) Fock term, iterated OPE and
irreducible two-pion exchange.
Medium insertions are systematically 
applied on all nucleon propagators, and
the relevant loop integrations yield results which can be written in analytic
form for all pieces. 

We now outline the leading contributions to the energy per particle $\bar{E}
(k_f)$. The kinetic energy including first order relativistic corrections is
\begin{equation}
\bar{E}_{kin} (k_f) = \frac{3 k^2_f}{10 M_N} \left( 1 - \frac{5 k^2_f}{28 M_N^2} 
\right).
\label{Ekin}
\end{equation}
Terms of order $k^6_f$ are already negligibly small. At least from this
perspective, nuclear matter is a non-relativistic system. 

The OPE Fock term (see the diagram on the left of Fig. \ref{VWfig1}) becomes
\begin{equation}
\bar{E}_{1 \pi} (k_f) = \frac{g^2_A m^3_{\pi}}{(4 \pi f_{\pi})^2} \left[F\left(
\frac{k_f}{m_{\pi}} \right) + \frac{m^2_{\pi}}{M_N^2} G\left(\frac{k_f}{m_{\pi}} \right) \right],
\label{Efock}
\end{equation}
where $F$ and $G$ are functions of the dimensionless variable $k_f /
m_{\pi}$. They are given explicitly in Ref. \cite{KFW1}. All finite parts of
iterated OPE and irreducible two-pion exchange are of the generic form
\begin{equation}
\bar{E}_{2 \pi} (k_f) = \frac{m^4_{\pi}}{(4 \pi f_{\pi})^4} \left[ g^4_A M_N
 H_4\left(\frac{k_f}{m_{\pi}} \right) + m_{\pi} H_5 \left(\frac{k_f}{m_{\pi}}
\right) \right],
\label{E2pi}
\end{equation}
with the functions $H_{4,5}$ again given explicitly in Ref.\cite{KFW1}. 
All power
divergences specific to cutoff regularization are summarized in the expression
\begin{equation}
\bar{E}_{\Lambda} (k_f) = \frac{\Lambda k^3_f}{(4 \pi f_{\pi})^4} [ - 10 g^4_A
M_N + (3 g^2_A + 1) (g^2_A - 1) \Lambda ],
\label{Elambda}
\end{equation}
where the attractive and dominant first term in the brackets arises from
iterated OPE. Note that this term, proportional to the density $\rho \sim k_f^3$,
can be generated in an equivalent mean-field approach by a
$NN$ contact interaction with appropriate coupling strength.The sum of the 
terms (\ref{Ekin})--(\ref{Elambda}) determines the energy per particle of 
nuclear matter within chiral dynamics, to the order shown in Fig.~\ref{VWfig1}.

\subsection{Nuclear matter equation of state}

\index{nuclear matter equation of state (EOS)}
A striking feature of the chiral dynamics approach is the simplicity of the
saturation mechanism for isospin-symmetric nuclear matter. Before turning to
the presentation of detailed results, it is instructive first to discuss the
situation in the exact chiral limit, $m_{\pi} = 0$. The basic saturation
mechanism can already be demonstrated by truncating the one- and two-pion
exchange diagrams at order $k^4_f$. We can make straightforward contact with
the parametrization (\ref{simple_eos}) of the energy per particle and identify the
coefficients $\alpha$ and $\beta$ of the $k^3_f$ and $k^4_f$ terms,
respectively. The result for $\alpha$ in the chiral limit is:
\begin{equation}
\alpha = \left( \frac{g_{\pi N}}{4 \pi} \right)^2
\left[\frac{10 \Lambda}{M_N} \left( \frac{g_{\pi N}}{4 \pi} \right)^2
 - 1\right] ,
\end{equation}
where we have neglected the small correction proportional to $\Lambda^2$ in
Eq.(\ref{Elambda}). The strongly attractive leading term in Eq.(\ref{Elambda}) is accompanied 
by the (weakly repulsive) one-pion exchange Fock term.
\begin{figure}[t]
\begin{center}
\parbox{8.5cm}{
\includegraphics[width=8.5cm,clip=true]{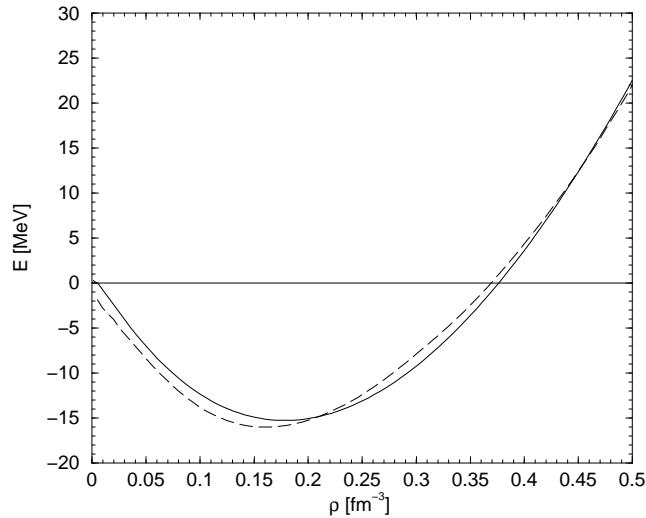}}
\end{center}
\caption{Energy per particle, $\bar{E} (k_f)$, of symmetric nuclear matter
derived from chiral one- and two-pion exchange (solid line) \cite{KFW1}. The
cutoff scale is $\Lambda = 646 \, MeV$. The dashed line is the result of Ref. 
\cite{FP.81}.}
\label{eos1}
\end{figure}
The $k^3_f$-contribution to $\bar{E} (k_f)$ would lead to collapse of the 
many-body
system. The stabilizing $k^4_f$-term is controlled by the coefficient
(calculated again in the chiral limit)
\begin{equation}
\beta = \frac{3}{70} \left( \frac{g_{\pi N}}{4 \pi} \right)^4 ( 4 \pi^2 + 237 -
24 \ln 2) - \frac{3}{56} = 13.55,
\label{beta}
\end{equation}
a unique and parameterfree result to this order. The two-pion exchange
dynamics in combination with the Pauli exclusion principle produces repulsion 
of just the right magnitude to achieve saturation:
the result, Eq.(\ref{beta}), is within $10 \%$ of the empirical $\beta =
12.2$. Adjustment of the short-distance scale $\Lambda$ between $0.5$ and 
$0.6$ GeV
easily leads to a stable minimum of $\bar{E} (k_f)$ in the proper range of
density and binding energy.

The full 3-loop  chiral dynamics result for $\bar{E} (k_f)$ in symmetric
nuclear matter, using $m_{\pi} = 135$ MeV (the neutral pion mass), is shown
in Fig.\ref{eos1} together with a realistic many-body calculation \cite{FP.81}. The outcome is
remarkable: with one single parameter $\Lambda = 0.65$ GeV fixed to the
value $\bar{E}_0 = -15.3$ MeV at equilibrium, perturbative pion
dynamics alone produces an equation of state which follows that of
much more sophisticated calculations up to about three times the
density of nuclear matter. The predicted compression modulus is $K_0
= 255$ MeV, well in line with the "empirical" $K_0 = (250 \pm 25)$ MeV.
The basic mechanisms behind binding and saturation in this approach are thus
identified as follows. The exchange of two pions between nucleons (truncated at the momentum scale $\Lambda$) drives the
attraction\footnote{In a qualitative (but not strict) analogy, these mechanisms are reminiscent 
of van der Waals forces.}
proportional to $k_f^3$ in the energy per particle. The Pauli blocking of intermediate nucleon
states in these $2\pi$ exchange processes produces the (repulsive) stabilizing $k_f^4$ term 
in $E/A$.
  
The calculations just described have also been extended to finite temperature
\cite{KFW1}. The resulting pressure as function of density shows the van der
Waals - like behaviour of the liquid - gas phase transition, with a critical
temperature $T_c\simeq 25$ MeV, slightly larger than the value around 
$18$ MeV commonly quoted as the ``empirical'' one.

\subsection{Asymmetry energy}

\index{asymmetry energy}
The specific isospin dependence of two-pion exchange should have its distinct
influence on the behaviour of asymmetric nuclear matter, with increasing excess
of neutrons over protons. We introduce as usual the asymmetry parameter
$\delta = (\rho_n - \rho_p) / \rho = (N-Z) / (N+Z)$, keeping the total density
$\rho = \rho_n + \rho_p = 2 k^3_f/ 3 \pi^2$ constant. The proton and neutron
densities are $\rho_{p,n} = (k_f^{p,n})^3 / 3 \pi^2$ in terms of the corresponding
Fermi momenta, $k_f^p$ and  $k_f^n$. 
Without change of any input, the
asymmetry energy $A (k_f)$ defined by
\begin{equation}
\bar{E}_{as} (k_f^p , k_f^n ) = \bar{E} (k_f) + \delta^2 A (k_f) +...
\end{equation}
has been calculated \cite{KFW1}.
The result at nuclear matter density is $A_0 = A(k_{f0}) = 33.8$
MeV. This is in very good agreement with the empirical value $A_0 = (30\pm 4)
$ 
MeV derived from extensive fits to nuclide masses \cite{atommass}.

Extrapolations to higher density work roughly up to $\rho \simeq 1.5 \,
\rho_0$. At still higher densities, there are indications that non-trivial
isospin dependence at shorter distances, beyond one- and two-pion exchange,
starts to play a role. A
similar statement holds for pure neutron matter which comes out
properly unbound, but
its predicted equation of state starts to deviate from that of realistic
many-body calculations at neutron densities larger than $0.2$ fm$^{-3}$.

\subsection{Nuclear mean field from chiral dynamics}

The in-medium three-loop calculation of the energy per particle defines the
(momentum dependent) self-energy of a single nucleon in nuclear matter up to
two-loop order. The real part of the resulting single particle potential in
isospin-symmetric matter at the saturation point, for a nucleon with zero
momentum, comes out as \cite{KFW2}
\begin{equation}
U (p = 0, k_{f0}) = -53.2 \, MeV,
\end{equation}
using exactly the same one- and two-pion exchange input that has led to the
solid curve in Fig. \ref{eos1}. The momentum dependence of $U (p, k_{f0})$ can be 
rephrased in
terms of an average effective nucleon mass $M^* \simeq 0.8 \, M_N$ at nuclear
matter density, and the imaginary part of the potential for a nucleon-hole at
the bottom of the Fermi sea is predicted to be about $30$ MeV. While these
numbers are remarkably close to the empirically deduced ones, it must nevertheless
be emphasised that the detailed momentum dependence of the single particle
potential, as obtained in this approach, is not yet satisfactory. 

\subsection{Intermediate summary and a working hypothesis}

Explicit pion dynamics originating from the spontaneously broken chiral 
symmetry
of QCD is clearly an important aspect of the nuclear many-body problem. 
In-medium
chiral perturbation theory, with one single cutoff scale $\Lambda \simeq 0.65$
GeV introduced to regularize the few divergent parts associated with
two-pion exchange, gives realistic binding and saturation of nuclear matter
already at three-loop order. At the same time it gives very good values for the
compression modulus and the asymmetry energy. These are non-trivial
observations, considering that it all works with only one adjustable parameter
encoding short-distance dynamics which remains unresolved at the Fermi
momenta $k_f \ll M_N \sim 4\pi f_{\pi}$ characteristic of equilibrium
nuclear matter. 

In view of the relevant scales in nuclear matter, the importance of
explicit pion degrees of freedom does not come unexpected. Many of the
existing models ignore pions, however. They introduce purely
phenomenological scalar fields with non-linear couplings and freely adjustable
parameters in order to simulate two-pion exchange effects.

Now, in order to establish contacts with the widely used and successful 
relativistic nuclear mean field phenomenology, the following working 
hypothesis suggests itself as a guide for the next necessary step. 
Assume that the nuclear matter ground state represents
a "shifted" QCD vacuum characterized by strong vector ($V$) and scalar ($S$)
condensate fields acting on the nucleons, with $V \simeq -S \simeq 0.3$ GeV
as suggested e.~g. by in-medium QCD sum rules \cite{CFG.91,DL.90,Jin.94}. 
Such a scenario would
not produce binding all by itself. Binding and saturation would instead result 
from pionic (chiral) fluctuations as outlined previously. 
The strong condensate fields
would, on the other hand, generate the large spin-orbit splitting in finite 
nuclei. This is the conceptual framework that we now apply when turning to
finite nuclei. 

\section{\label{PC_model}Relativistic nuclear model constrained by QCD and chiral symmetry}

\index{quantum hadrodynamics (QHD)}
At this point it is appropriate to recall in greater detail the relativistic nuclear 
phenomenology that goes under the name of 
Quantum Hadrodynamics (QHD)~\cite{SW.97}: a framework of Lorentz-covariant
meson-baryon effective field theories.
In the mean-field approximation, such an approach is equivalent 
to a model with local four-point
interactions between nucleons~\cite{Mad,Rusnak:1997dj,Burvenich:2001rh}. 
Models based on QHD have been successfully 
applied to describe a broad range of nuclear phenomena, 
from light nuclei to superheavy elements (see Ref.~\cite{Rin.96} for a
recent review, and references therein). This framework has also been 
extended to studies of 
the structure of exotic nuclei with extreme isospin and close to the particle drip lines.

The effective hadronic Lagrangians of QHD and related models incorporate known long-range interactions constrained by symmetries and a set of generic short-range interactions. These models
are consistent with fundamental symmetries: Lorentz invariance, parity invariance, electromagnetic gauge invariance, isospin and with the chiral symmetry of QCD \cite{SW.97,FS.00,FTS}.  
The most successful QHD models are, however, 
purely phenomenological, with parameters adjusted to reproduce 
the nuclear matter equation of state and global properties 
of spherical closed-shell nuclei. QHD calculations do not include
pions explicitly, whereas we have just demonstrated that pions, as Goldstone bosons 
of spontaneously broken chiral symmetry, play an important role in the nuclear many-body
problem. Two-pion exchange effects are supposedly incorporated as part of
the strong scalar-isoscalar field of QHD models, but in an ad-hoc manner
without detailed reference to the underlying $\pi\pi NN$ dynamics.
QCD symmetries do constrain the effective QHD Lagrangians by restricting the form of possible 
interaction terms. However, the empirical data set of bulk 
and single-particle properties of finite nuclei can only determine 
six or seven parameters in the general expansion of the effective 
Lagrangian in powers of the fields 
and their derivatives \cite{FS.00b}. Such a general expansion can be 
controlled by the ``naive dimensional analysis" 
(NDA)~\cite{Rusnak:1997dj,Burvenich:2001rh,Man83,FML.96}. NDA tests the 
coefficients of the expansion for ``naturalness", i.e. this method 
controls the orders of magnitude of the coupling constants. NDA can 
exclude some interaction terms because their couplings would be 
``unnatural", but it cannot determine the model parameters at the 
level of accuracy required for a quantitative analysis of nuclear structure data. 

Our approach is similar in spirit but proceeds with a different strategy, 
imposing as many low-energy QCD constraints as possible in order to minimize the number of freely adjustable parameters. It is based on the conjectures already presented in the Introduction, namely:\\

that nuclear binding and saturation arise dominantly 
from chiral (pionic) fluctuations in combination with Pauli blocking effects,
as described in the previous section;

that these pionic fluctuations are superimposed on individually strong scalar and vector 
fields of approximately equal magnitude and opposite sign, 
originating from density-dependent changes of the QCD vacuum condensates.\\ 

Our aim is thus to study the interplay between condensate
background fields and perturbative chiral fluctuations, both rooted in the
spontaneous symmetry breaking pattern of QCD, in forming nuclei. We will
demonstrate how this scenario works at large. Whereas in first
approximation the condensate fields do not play a significant role for the 
saturation mechanism, they are essential for the description of ground states of finite nuclei and, 
in particular, the spin-orbit force.

\subsection{Point-coupling model: density dependent contact interactions}

\index{point-coupling model}
A suitable framework to address these questions, both for nuclear matter
and for finite nuclei, is a relativistic point coupling model with 
density dependent $NN$ contact interactions. The explicit density dependence
of these couplings reflects effects ``beyond mean field'' such as they
arise from the Pauli corrections to two-pion exchange fluctuations discussed
previously. The mapping of the $k_f$ expansions of chiral dynamics, described 
in Section 3, into the form of equivalent point couplings depending on 
fractional powers of $\rho$, is performed by equating the nucleon 
self-energies resulting from both schemes.

\subsubsection{A. Lagrangian} 

The relativistic point-coupling Lagrangian is 
built from basic density and current operator blocks bilinear in the Dirac 
spinor field $\Psi$ of the nucleon\footnote{From here on we denote the nucleon field by $\Psi$,
dropping the index $N$. Similarly, the nucleon mass is denoted $M$ without further index specification.} : 
\begin{equation} 
  ( {\bar{\Psi}} {\mathcal O}_\tau \Gamma  {\Psi}) 
  \quad,\quad 
  {\mathcal O}_\tau\in\{ {1},\tau_i\} 
  \quad,\quad 
  \Gamma\in\{1,\gamma_\mu,\gamma_5,\gamma_5\gamma_\mu,\sigma_{\mu\nu}\}\; . 
\end{equation} 
Here $\tau_i$ are the isospin matrices and 
$\Gamma$ generically denotes the Dirac matrices. 
The interaction terms of the Lagrangian are products of these 
blocks to a given order. In principle, a general effective Lagrangian 
can be written as a power series in the point couplings 
$( {\bar{\Psi}} {\mathcal O}_\tau \Gamma  {\Psi})$ and their derivatives. 
In practice, properties of symmetric and asymmetric 
nuclear matter, as well as empirical ground state properties of finite 
nuclei, constrain only the isoscalar-scalar, the isoscalar-vector, 
the isovector-vector, and to a certain extent the isovector-scalar 
channels. 
\par 
The model that we consider here 
includes the following four-fermion interaction vertices: 
\begin{center} 
\begin{tabular}{ll} 
 isoscalar-scalar:   &   $(\bar\Psi\Psi)^2$\\ 
 isoscalar-vector:   &   $(\bar\Psi\gamma_\mu\Psi)(\bar\Psi\gamma^\mu\Psi)$\\ 
 isovector-scalar:  &  $(\bar\Psi\vec\tau\Psi)\cdot(\bar\Psi\vec\tau\Psi)$\\ 
 isovector-vector:   &   $(\bar\Psi\vec\tau\gamma_\mu\Psi) 
                         \cdot(\bar\Psi\vec\tau\gamma^\mu\Psi)$ .\\ 
\end{tabular} 
\end{center} 
Vectors in isospin space are denoted by boldface symbols.
The model is defined by the Lagrangian density 
\begin{equation} 
\mathcal{L} = \mathcal{L}_{\rm free} + \mathcal{L}_{\rm 4f} 
  + \mathcal{L}_{\rm der} + \mathcal{L}_{\rm em}, 
\label{Lag} 
\end{equation} 
with the four terms specified as follows: 
\begin{eqnarray} 
\label{Lag2} 
\mathcal{L}_{\rm free} & = &\bar{\Psi}(i\gamma_{\mu}\partial^{\mu} -M)\Psi \; ,\\ 
\label{Lag3} 
\mathcal{L}_{\rm 4f} & = & 
   - \frac{1}{2}~G_{S}(\hat{\rho}) (\bar{\Psi}\Psi)(\bar{\Psi}\Psi)  \nonumber\\
   & ~ & -\frac{1}{2}~G_{V}(\hat{\rho})(\bar{\Psi}\gamma_{\mu}\Psi) 
   (\bar{\Psi}\gamma^{\mu}\Psi) \nonumber\\ 
   & ~ & - \frac{1}{2}~G_{TS}(\hat{\rho}) 
   (\bar{\Psi}\vec{\tau}\Psi) \cdot (\bar{\Psi} \vec{\tau} \Psi) \nonumber\\ 
   & ~ & - \frac{1}{2}~G_{TV}(\hat{\rho})(\bar{\Psi}\vec{\tau} 
   \gamma_{\mu}\Psi)\cdot (\bar{\Psi}\vec{\tau} \gamma^{\mu}\Psi) \; ,\\ 
\label{Lag4} 
\mathcal{L}_{\rm der} & = & -\frac{1}{2}~D_{S}(\hat{\rho}) (\partial_{\nu} 
  \bar{\Psi}\Psi)(\partial^{\nu}\bar{\Psi}\Psi)\nonumber\\ 
 & ~ & - \frac{1}{2}~D_{V}(\hat{\rho}) 
(\partial_{\nu}\bar{\Psi}\gamma_{\mu} 
  \Psi)(\partial^{\nu}\bar{\Psi}\gamma^{\mu}\Psi)\nonumber\\ 
 & ~ & - \frac{1}{2}~D_{TS}(\hat{\rho}) 
(\partial_{\nu}\bar{\Psi}\vec{\tau} 
  \Psi) \cdot (\partial^{\nu}\bar{\Psi}\vec{\tau}\Psi) \nonumber\\
 & ~ & - \frac{1}{2}~D_{TV}(\hat{\rho}) 
(\partial_{\nu}\bar{\Psi}\vec{\tau} 
  \gamma_{\mu}\Psi) \cdot (\partial^{\nu}\bar{\Psi}\vec{\tau} 
  \gamma^{\mu}\Psi)\; , \\
\label{Lag5} 
\mathcal{L}_{\rm em} & = & 
+eA^{\mu}\bar{\Psi}\frac{1+\tau_3}{2}\gamma_{\mu}\Psi 
  -\frac{1}{4} F_{\mu\nu}F^{\mu\nu} \; . 
\end{eqnarray} 
While the Lagrangian (\ref{Lag})-(\ref{Lag5}) is going to be formally used 
in the mean-field approximation, fluctuations beyond mean field 
are nevertheless encoded 
in the density-dependent couplings $G_i(\hat{\rho})$ and $D_i(\hat{\rho})$, 
to be specified in detail later. 
When applied to finite nuclei the model 
must also include the coupling $\mathcal{L}_{\rm em}$
of the protons to the electromagnetic field, 
and the derivative terms in $\mathcal{L}_{\rm der}$, in addition to the free 
nucleon Lagrangian $\mathcal{L}_{\rm free}$ 
and the interaction terms in $\mathcal{L}_{\rm 4f}$. The 
derivative terms are built from the blocks 
$\partial_\nu(\bar\Psi\tau_i\Gamma_j\Psi)$ 
(the derivative is understood to act on both $\bar\Psi$ and $\Psi$). 
One can, of course, construct many more derivative terms
of higher orders. 
In $\mathcal{L}_{\rm der}$ we include only those terms which 
are direct counterparts of the second order interaction terms 
incorporated in $\mathcal{L}_{\rm 4f}$. In this way we take into 
account leading effects of finite range 
interactions that are important for a quantitative fine-tuning
of nuclear properties.   

Our model Lagrangian formally resembles the ones used in 
the standard relativistic mean-field point-coupling models of 
Refs.~\cite{Mad,Rusnak:1997dj,Burvenich:2001rh,FML.96,Manakos:wu}. 
The underlying dynamics is, however, quite different. 
The parameters of the interaction terms in the standard 
point-coupling models have constant values adjusted to reproduce 
the nuclear matter equation of state and a set of global 
properties of spherical closed-shell nuclei. In order to 
describe properties of finite nuclei on a quantitative level, 
these models include also some higher order interaction terms, 
such as six-fermion vertices $(\bar\Psi\Psi)^3$, and 
eight-fermion vertices $(\bar\Psi\Psi)^4$ and 
$[(\bar\Psi\gamma_\mu\Psi)(\bar\Psi\gamma^\mu\Psi)]^2$. 
Our approach includes only 
second order interaction terms, but with coupling strengths that are functions 
of the nucleon density operator $\hat{\rho}$. Their functional dependence 
will be determined by matching the nucleon self-energies resulting from in-medium chiral perturbation theory  and finite-density QCD sum rules. 

Medium dependent vertex functions have also been considered 
in so-called density dependent relativistic hadron field (DDRH) 
models. The density dependent meson-nucleon vertex functions are 
determined either 
by mapping the nuclear matter Dirac-Brueckner nucleon self-energies 
in the local density approximation~\cite{FLW.95,JL.98,HKL.01}, or 
they are adjusted 
to reproduce properties of symmetric and asymmetric nuclear matter 
and finite nuclei~\cite{TW.99,Niksic:2002yp}. In practical applications 
of the DDRH models the meson-nucleon couplings are assumed to be 
functions of the baryon density $\Psi^\dagger \Psi$. 
In a relativistic framework the couplings can also depend on 
the scalar density $\bar{\Psi} \Psi$. Nevertheless, expanding in $\Psi^\dagger \Psi$
is the natural choice, for several reasons. The baryon density is connected to the conserved
baryon number, unlike the scalar density for which no conservation law exists. The scalar density 
is a dynamical quantity, to be determined self-consistently by the equations of motion, and expandable in powers of the Fermi momentum. For the meson exchange 
models it has been shown that the dependence on baryon density alone provides a more direct
relation between the self-energies of the density-dependent 
hadron field theory and the Dirac-Brueckner microscopic 
self-energies~\cite{HKL.01}. Moreover, the pion-exchange 
contributions to the nucleon self-energy, as calculated using in-medium chiral perturbation theory, are 
directly given as expansions in powers of the Fermi momentum. Following these considerations, we 
express the coupling strengths of the interaction terms 
as functions of the baryon density, represented by the operator 
$\hat{\rho}=\Psi^\dagger \Psi$ in the rest frame of the many-body system. 

\subsubsection{B. Equation of motion and nucleon self-energies}  

The single-nucleon Dirac equation is derived by variation of the 
Lagrangian (\ref{Lag}) with respect to $\bar{\Psi}$: 
\begin{equation} 
\label{Dirac} 
[\gamma_{\mu}(i\partial^{\mu} - V^\mu) - 
   (M + S) ]\Psi = 0\; , 
\end{equation} 
where 
\begin{equation} 
   V^\mu = \Sigma^{\mu} + 
   \vec{\tau} \cdot \vec{\Sigma}^{\mu}_{T} 
   +\Sigma_{r}^{\mu} 
   +\vec{\tau} \cdot \vec{\Sigma}_{rT}^{\mu} 
\end{equation}
and
\begin{equation}
   S = \Sigma_S + \vec{\tau} \cdot \vec{\Sigma}_{TS} + \Sigma_{rS} 
   +\vec{\tau} \cdot \vec{\Sigma}_{rTS} \; , 
\end{equation} 
with the nucleon self-energies defined by the following relations 
\begin{eqnarray} 
\label{self1} 
\Sigma^{\mu} & = & (G_V - D_V \Box) j^{\mu} - 
   eA^{\mu}\frac{1+\tau_3}{2}\\ 
\label{self2} 
\vec{\Sigma}^{\mu}_{T} & = & (G_{TV} - D_{TV} \Box ) \vec{j}_{T}^{\mu}\\ 
\label{self3} 
\Sigma_S & = & ( G_S - D_S \Box ) (\bar{\Psi} \Psi)\\ 
\label{self4} 
\vec{\Sigma}_{TS} & = & ( G_{TS} - D_{TS} \Box ) (\bar{\Psi} \vec{\tau} \Psi)\\ 
\label{self5} 
\Sigma_{rS} &  = & - \frac{\partial D_S}{\partial \hat{\rho}} (\partial_{\nu} 
   j^{\mu}) u_{\mu} (\partial^{\nu} (\bar{\Psi} \Psi))\\ 
\label{self6} 
\vec{\Sigma}_{rTS} &  = & - \frac{\partial D_{TS}}{\partial \hat{\rho}} 
(\partial_{\nu} 
   j^{\mu}) u_{\mu} (\partial^{\nu} (\bar{\Psi} \vec{\tau} \Psi))\\ 
\label{self7} 
\vec{\Sigma}_{rT}^{\mu} & = & -\frac{\partial D_{TV}}{\partial \hat{\rho}} 
   (\partial_{\nu} j^{\alpha}) u_{\alpha} (\partial^{\nu} 
   \vec{j}_{T}^{\mu})\\ 
\label{self8} 
\Sigma_r^{\mu} & = & u^\mu \left( \frac{1}{2}   
   \frac{\partial G_S}{\partial \hat{\rho}} (\bar{\Psi} \Psi) 
   (\bar{\Psi} \Psi) + \frac{1}{2}
   \frac{\partial D_S}{\partial \hat{\rho}} (\partial^{\nu} (\bar{\Psi} \Psi)) 
   (\partial_{\nu} (\bar{\Psi} \Psi))\right. \nonumber\\ 
   & ~ & + \frac{1}{2}   
   \frac{\partial G_{TS}}{\partial \hat{\rho}} (\bar{\Psi} \vec{\tau} \Psi) \cdot
   (\bar{\Psi} \vec{\tau} \Psi) + \frac{1}{2}   
   \frac{\partial D_{TS}}{\partial \hat{\rho}} (\partial^{\nu} (\bar{\Psi} 
   \vec{\tau} \Psi)) \cdot (\partial_{\nu} (\bar{\Psi} \vec{\tau} \Psi)) \nonumber\\ 
 & ~ & + \frac{1}{2} \frac{\partial G_V}{\partial \hat{\rho}} j^{\nu} 
   j_{\nu} + \frac{1}{2} \frac{\partial D_V} 
   {\partial \hat{\rho}}(\partial_{\nu} 
   j_{\alpha})(\partial^{\nu} j^{\alpha})\nonumber\\ 
 & ~ & \left. + \frac{1}{2} \frac{\partial G_{TV}} 
   {\partial \hat{\rho}} \vec{j}_{T}^{\nu} \cdot \vec{j}_{T\nu}^{~} + 
   \frac{1}{2} \frac{\partial D_{TV}}{\partial \hat{\rho}} 
   (\partial_{\nu} \vec{j}_{T\alpha}^{~}) \cdot (\partial^{\nu} \vec{j}_{T}^{\alpha}) 
   \right) \nonumber\\ 
 & ~ & - \frac{\partial D_V}{\partial \hat{\rho}} 
   (\partial_{\nu} j_{\alpha}) u^{\alpha} (\partial^{\nu}j^{\mu}) \; . 
\end{eqnarray} 
\index{rearrangement self-energies} 
The nucleon isoscalar and isovector currents read 
\begin{eqnarray} 
j^{\mu} & = & \bar{\Psi} \gamma^{\mu} \Psi \; ,\\ 
\vec{j}_T^{\mu} & = & \bar{\Psi} \vec{\tau} \gamma^{\mu} \Psi \; ,   
\end{eqnarray} 
respectively. We write $\hat{\rho} u^{\mu} = j^{\mu}$ 
and the four-velocity $u^{\mu}$ is defined 
as $(1-{\bf v}^2)^{-1/2}(1,{\bf v})$ 
where ${\bf v}$ is the three-velocity vector (${\bf v}=0$ 
in the rest-frame of the nuclear system). 
In addition to the isoscalar-vector $\Sigma^{\mu}$, 
isoscalar-scalar $\Sigma_S$, isovector-vector $\vec{\Sigma}^{\mu}_{T}$ 
and isovector-scalar $\vec{\Sigma}_{TS}$ 
self-energies, the density dependence of the vertex functions 
produces the {\it rearrangement} contributions $\Sigma_{rS}$, $\Sigma_r^{\mu}$, 
$\vec{\Sigma}_{rTS}$ and $\vec{\Sigma}_{rT}^{\mu}$. 
The rearrangement terms result from the variation of the vertex 
functionals with respect to the baryon fields in the density operator $\hat{\rho}$ 
(which coincides with the baryon density in the nuclear matter rest-frame). 
For a model with density dependent couplings, the inclusion 
of the rearrangement self-energies 
is essential for energy-momentum conservation 
$\partial_\mu T^{\mu \nu} = 0$,
and thermodynamical consistency 
${\rho^2 \frac{\partial}{\partial \rho}
\left( \frac{\varepsilon}{\rho} \right) = \frac{1}{3}
\sum_{i=1}^3 T^{ii} }$ 
(i.e. for the pressure equation derived from 
the thermodynamic definition and from the 
energy-momentum tensor)~\cite{FLW.95,TW.99}. 

When applied to nuclear matter or ground-state properties of 
finite nuclei, the point-coupling model is understood to be used in the 
mean-field approximation. 
The ground state of a nucleus with A nucleons 
is the product of the lowest occupied single-nucleon 
self-consistent stationary solutions of the Dirac equation (\ref{Dirac}). 
The ground state energy is the sum of the single-nucleon energies plus 
a functional of the scalar density, 
\begin{equation} 
\rho_s = \sum\limits_{k=1}^{A} \bar \psi_k \psi_k \,, 
\end{equation} 
and of the baryon density of the nucleons, 
\begin{equation} 
\rho = \sum\limits_{k=1}^{A} \psi^{\dagger}_k \psi_k \,, 
\end{equation} 
where the sums run over occupied positive-energy single-nucleon 
states $|k\rangle$ with wave functions $\psi_k$. 

The density dependence of the coupling strengths which determine 
the self-energies 
(\ref{self1}-\ref{self8}), will be constrained by in-medium QCD sum rules 
and chiral pion dynamics, to be specified in detail in the following sections  

\subsection{Nuclear matter, part II}

\index{nuclear matter equation of state (EOS)}
In translationally invariant infinite nuclear matter 
all terms involving the derivative couplings (\ref{Lag4}) drop out and 
the single-nucleon Dirac equation reads 
\begin{equation} 
[\gamma_{\mu}(i\partial^{\mu} - \Sigma^{\mu} - \Sigma^{\mu}_r - \vec{\tau} \cdot 
\vec{\Sigma}^{\mu}_{T}) 
   - (M + \Sigma_S + \vec{\tau} \cdot \vec{\Sigma}_{TS})]\Psi = 0 \; , 
\label{Dirac-NM} 
\end{equation} 
where the self-energies are
\begin{eqnarray} 
\label{selfnm1} 
\Sigma^{\mu} & = & G_V j^{\mu} \; ,\\ 
\label{selfnm2} 
\Sigma^{\mu}_r & = & u^\mu \left[ \frac{1}{2} \frac{\partial G_S} 
   {\partial \hat{\rho}} (\bar{\Psi} \Psi) (\bar{\Psi} \Psi) + 
   \frac{1}{2} \frac{\partial G_V}{\partial \hat{\rho}} 
   j^{\nu}j_{\nu} \right. \nonumber\\ 
   & ~ &  \left. + \frac{1}{2} \frac{\partial G_{TS}} 
   {\partial \hat{\rho}} (\bar{\Psi} \vec{\tau} \Psi) \cdot (\bar{\Psi} 
   \vec{\tau} \Psi) + 
   \frac{1}{2} \frac{\partial G_{TV}}{\partial \hat{\rho}} 
   \vec{j}_{T}^{\nu} \cdot \vec{j}_{T\nu}^{~} \right]\label{SE-SV} \; ,\\   
\label{selfnm3} 
\vec{\Sigma}^{\mu}_{T} & = & G_{TV} \vec{j}_{T}^{\mu}\label{SE-VV} \; ,\\ 
\label{selfnm4} 
\Sigma_S & = & G_S (\bar{\Psi} \Psi)\label{SE-SS} \; ,\\ 
\label{selfnm5} 
\vec{\Sigma}_{TS} & = & G_{TS} (\bar{\Psi}\vec{\tau} \Psi)\label{SE-VS} \; . 
\end{eqnarray} 
The total isoscalar vector self-energy includes the rearrangement contributions 
$\Sigma^{\mu}_r$. 
In nuclear matter at rest the spatial components of the four-currents vanish, 
and the densities are calculated by taking expectation values 
\begin{eqnarray} 
\rho_s & = & \langle \Phi|\bar{\Psi} \Psi|\Phi \rangle \; , \\ 
\rho & = & \langle \Phi|\bar{\Psi}\gamma^{0} \Psi|\Phi \rangle \; , \\ 
\rho_{s3} & = & \langle \Phi|\bar{\Psi} \tau_3 \Psi|\Phi \rangle \; , \\ 
\rho_{3} & = & \langle \Phi|\bar{\Psi} \tau_3 \gamma^{0} \Psi|\Phi \rangle \; , 
\end{eqnarray} 
where $|\Phi \rangle$ is the nuclear matter ground state. 
The energy density ${\cal E}$ and the pressure $P$ are derived from the 
energy-momentum tensor $T^{\mu\nu}$ as 
\begin{eqnarray} 
{\cal E} & = & {\cal E}_{kin}^n + {\cal E}_{kin}^p 
   - \frac{1}{2}G_S \rho_s^2 - \frac{1}{2}G_{TS} \rho_{s3}^2 
   + \frac{1}{2}G_V \rho^2 + \frac{1}{2}G_{TV} \rho_{3}^2 \; ,\label{epsilon}\\ 
 & ~ & \nonumber \\ 
P & = & \tilde{E}_p \rho_p + \tilde{E}_n \rho_n - {\cal E}_{kin}^p - 
   {\cal E}_{kin}^n 
   + \frac{1}{2} G_V \rho^2 + \frac{1}{2} G_{TV}\rho_{3}^2  
   + \frac{1}{2} G_S \rho_s^2 + \frac{1}{2} G_{TS} \rho_{s3}^2 \nonumber \\ 
 & ~ &
   + \frac{1}{2} \frac{\partial G_S} {\partial \rho} \rho_s^2 \rho + 
   \frac{1}{2} \frac{\partial G_V}{\partial \rho} \rho^3 + 
   \frac{1}{2} \frac{\partial G_{TV}}{\partial \rho} \rho_{3}^2 
   \rho + \frac{1}{2} \frac{\partial G_{TS}} {\partial \rho} 
   \rho_{s3}^2 \rho \label{pressure} \; . 
\end{eqnarray} 
The baryon density is related to the Fermi momentum $k_f$ in the usual way, 
\begin{equation} 
\rho_i = \frac{2}{(2\pi)^3} \int\limits_{|k|\le k^i_f} d^3 k  = 
    \frac{(k^i_f)^3}{3\pi^2} \; , 
\end{equation} 
where the index $i=\{p,n\}$ refers to protons and neutrons, respectively. 
The corresponding scalar densities are determined by the self-consistency 
relation 
\begin{equation} 
\rho_s^i = \frac{2}{(2\pi)^3} \int\limits_{|k|\le k^i_f} d^3 k 
    \frac{M^*_i}{\sqrt{k^2 
    + (M^*_i)^2}} 
    = \frac{M^*_i}{2\pi^2} \left[ k^i_f \tilde{E}_i - (M^*_i)^2 
    \ln \frac{k^i_f + \tilde{E}^i}{M^*_i} \right] \; , 
\end{equation} 
with the proton and neutron quasi-particle energies 
\begin{equation} 
\tilde{E}_i = \sqrt{(k^i_f)^2 + (M^*_i)^2} \; , 
\end{equation} 
and the effective nucleon masses 
\index{effective mass}
\begin{eqnarray} 
M^*_p &=& M + G_S \,\rho_s + G_{TS}\, \rho_{s3} \; ,\\ 
M^*_n &=& M + G_S \,\rho_s - G_{TS} \,\rho_{s3} \; . 
\end{eqnarray} 
The kinetic contributions to the energies of the protons and neutrons in 
nuclear matter are calculated from 
\begin{equation} 
{\cal E}^i_{kin} = \frac{2}{(2\pi)^3} \int\limits_{|k|\le k^i_f} 
    d^3 k \sqrt{k^2 + (M^*_i)^2} = 
    \frac{1}{4} [ 3 \tilde{E}_i \rho_i + M^*_i \rho_s^i ] \quad i=\{p,n\}\; . 
\end{equation} 
Note that {\it rearrangement} contributions 
appear explicitly in the expression for the pressure, whereas in the 
energy density, such contributions 
are implicit through the specific density dependence of $G_V(\hat{\rho})$.


\subsection{QCD constraints} 
 
We proceed now with a central theme of this section: establishing connections 
between the density-dependent point couplings in the Lagrangian (\ref{Lag})
and constraints from QCD. Two key features of low-energy, 
non perturbative QCD are at the origin 
of this discussion: 
the presence of a non-trivial vacuum characterized by strong 
condensates and the important role  of pionic fluctuations governing 
the low-energy, long wavelength dynamics according to the rules imposed by 
spontaneously broken chiral symmetry. 
Our basic conjecture is, consequently, 
that the nucleon isoscalar self-energies (\ref{selfnm1}) and 
(\ref{selfnm4}) arise primarily 
through changes in the quark condensate and in the quark density at 
finite baryon density, together with chiral (pionic) fluctuations 
induced by one- and two-pion exchange interactions. 

\subsubsection{A. In-medium QCD sum rules}

\index{QCD sum rules}
QCD sum rules combine dispersion relation methods with operator product
expansion techniques in the analysis of correlations functions. When
applied to the nucleon, the starting point is a suitably chosen 3-quark
operator $J_N(x)$ which carries nucleon quantum numbers. The correlation function 
\begin{equation}
\Pi(q^2) = i\int d^4x\,e^{iq\cdot x}\langle {\cal T}J_N(x)\bar{J}_N(0)
\rangle \nonumber
\end{equation}
has a pole structure which determines the nucleon mass in the 
presence of the (non-perturbative) QCD vacuum. That same correlation
function can be written in terms of the \tindex{operator product expansion}, 
\begin{equation}
\Pi(q^2) = \sum_n C_n(q^2)\langle \hat{O}_n\rangle \, , 
\label{OPE}
\end{equation}
with vacuum expectation values of local operators $\hat{O}_n$ constructed from
quark and gluon fields and organised according to their (mass) dimensions. 
The operators of lowest non-trivial dimension, $d=4$, are $m_q\,\bar{q}q$ and 
$G_{\mu\nu}G^{\mu\nu}$. The next higher ones include four-quark operators and
combinations of quark and gluon fields.

Now, for sufficiently large spacelike $Q^2 = -q^2 >0$, the coefficients
$C_n(q^2)$, called \tindex{Wilson coefficients}, decrease as inverse powers of $Q^2$
with each of the higher dimensional operators to which they are attached.
One can then apply QCD perturbation theory to calculate the $C_n$. The next
step is to match $\Pi(-Q^2)$ with its dispersion relation representation and
apply a differential operation, the so-called \tindex{Borel transform},
to improve convergence properties. When this matching is performed, 
the leading order result for the nucleon
mass in vacuum is Ioffe's formula (\ref{Io}).

Applications of QCD sum at finite baryon density $\rho$ proceed in an 
analogous way, except that now operators
such as $q^{\dagger}q$ also have non-vanishing ground state expectation values:
$\langle q^{\dagger}q \rangle_{\rho} =
3\rho/2$. In addition, the vacuum condensates vary with increasing density.
Thus in-medium QCD sum rules relate the changes of the 
scalar quark condensate and the quark density at 
finite baryon density, with the isoscalar scalar and vector self-energies 
of a nucleon in the nuclear medium. In leading order which should be valid at 
densities below and around saturated nuclear matter, the condensate 
part, $\Sigma_S^{(0)}$, of the scalar self-energy is expressed in terms of the density dependent 
chiral condensate as follows~\cite{CFG.91,DL.90,Jin.94}: 
\begin{equation} 
\Sigma^{(0)}_S = - \frac{8 \pi^2}{\Lambda_B^2} [ 
\langle \bar{q} q \rangle_\rho - \langle \bar{q} q \rangle_0 ] 
 = - \frac{8 \pi^2}{\Lambda_B^2}~\frac{\sigma_N}{m_u +m_d} \rho_s \; ,
\end{equation} 
with $\rho_s = \langle \bar{\Psi}\Psi \rangle$. We have used Eq.(\ref{mediumcond}) together with
the GOR relation (\ref{GOR}). The chiral vacuum condensate $\langle \bar{q} q \rangle_0$ as a measure
of spontaneous chiral symmetry breaking in QCD is given in Eq.(\ref{cond}). The difference between the vacuum condensate $\langle \bar{q} q \rangle_0$ 
and the one at finite density involves the nucleon sigma term, 
$\sigma_N \sim \langle N| m_q \bar{q} q |N \rangle$, to this order. 
The Borel mass scale $\Lambda_B \approx 1$GeV roughly separates 
perturbative and non-perturbative domains in the QCD sum rule 
analysis. 

To the same order in the condensates with lowest dimension, 
the resulting time component of the isoscalar vector self-energy is 
\begin{equation} 
\Sigma_V^{(0)} = \frac{64 \pi^2}{3 \Lambda_B^2} \langle q^\dagger q \rangle_\rho 
= \frac{32 \pi^2}{\Lambda_B^2} \rho \; .
\end{equation}  
It reflects the repulsive density-density correlations associated with the time component
of the quark baryon current, $\bar{q}\gamma^{\mu}q$. 
Note that, as pointed out in Ref.\cite{CFG.91}, the ratio 
\begin{equation} 
\label{ratio} 
\frac{\Sigma_S^{(0)}}{\Sigma_V^{(0)}} = - \frac{\sigma_N}{4(m_u +m_d)} 
\frac{\rho_s}{\rho} 
\end{equation} 
is approximately equal to  $-1$ for typical values of 
the nucleon sigma term $\sigma_N$ and the current quark masses 
$m_{u,d}$, and around nuclear matter saturation density where $\rho_s \simeq \rho$ 
(as an example, take $\sigma_N \simeq  50$ MeV and $m_u +m_d \simeq  
12$ MeV at a scale of 1 GeV). 

Identifying the free nucleon mass at $\rho=0$ according to Ioffe's formula (\ref{Io}), 
$M = - \frac{8\pi^2}{\Lambda_B^2} \langle \bar{q} q \rangle_0$, one finds 
\begin{equation} 
\label{back1} 
\Sigma_S^{(0)} (\rho) = M^*(\rho) -M = 
- \frac{\sigma_N M}{m_\pi^2 f_\pi^2} \rho_s 
\end{equation} 
and 
\begin{equation} 
\label{back2} 
\Sigma_V^{(0)} (\rho) = \frac{4 (m_u + m_d)M}{m_\pi^2f_\pi^2} \rho \; , 
\end{equation} 
with the quark masses $m_{u,d}$ to be taken at a renormalization 
scale $\mu \simeq \Lambda_B \simeq 4\pi f_\pi \simeq 1$ GeV. 
\par
Given these self-energies arising from the condensate background, 
the corresponding equivalent point-coupling strengths $G_{S,V}^{(0)}$ 
are simply identified through the relations 
\begin{equation} 
\label{sigmaS} 
\Sigma_S^{(0)} = G_S^{(0)} \rho_s
\end{equation} 
and 
\begin{equation} 
\label{sigmaV} 
\Sigma_V^{(0)} = G_V^{(0)} \rho \;  
\end{equation}
in mean field approximation. To this order, the couplings $G_{S,V}^{(0)}$ are density independent. 
Eq.(\ref{back1}) implies (identifying $M$ with the free nucleon mass): 
\begin{equation} 
\Sigma_S^{(0)} \simeq -350~{\rm MeV}~\frac{\sigma_N} 
{50~{\rm MeV}} \,{\rho_s\over \rho_0} \; .
\end{equation}
Evidently, the leading-order in-medium change of the chiral condensate
is a source of a strong, attractive scalar field which acts
on the nucleon in such a way as to reduce its mass in nuclear matter
by more than $1/3$ of its vacuum value as already anticipated in Eq.(\ref{lin}).
Using (\ref{sigmaS}) one estimates: 
\begin{equation} 
\label{estimate} 
G_S^{(0)} \simeq  -11~{\rm fm}^2~\frac{\sigma_N}{50~{\rm MeV}} \quad {\rm at} 
\,\,\,\rho_s \simeq \rho_0 = 0.16~{\rm fm}^{-3}. 
\end{equation} 
A correspondingly smaller value of $G_{S}^{(0)}$ would result if 
only a fraction of the nucleon mass $M$ is associated with the chiral 
condensate, leaving room for non-leading 
contributions from higher dimensional condensates. 
The QCD sum rules suggest that the ratio of the condensate scalar and 
vector self-energies is close to $-1$, so one expects roughly 
$G_V^{(0)} \simeq - G_S^{(0)}$. 

One should note of course that the QCD sum rule constraints 
implied by Eqs.(\ref{back1}, \ref{back2}) and by the ratio (\ref{ratio}) 
are not very accurate at a quantitative level. 
The leading-order Ioffe formula on which Eq.(\ref{back1}) relies has corrections 
from condensates of higher dimension which are not well under control. 
The estimated error in the ratio $\Sigma_S^{(0)}/\Sigma_V^{(0)}\simeq -1$ 
is about 20\%, given the uncertainties in the values of $\sigma_N$ and $m_u+m_d$. 
Nevertheless, the constraints implied by Eq.(\ref{ratio}) give 
important hints for further orientation. 

\subsubsection{B. Pionic fluctuations: self-energies from in-medium chiral perturbation theory}

The second essential ingredient is the contribution to the nucleon self-energies 
from chiral fluctuations related to pion-exchange processes 
(primarily two-pion exchange interactions, 
with small corrections from one-pion-exchange Fock terms). 
The corresponding density dependent point-coupling strengths, denoted by $G^{(\pi)}_{S,V}(\rho)$ etc., 
are calculated using in-medium chiral perturbation theory.. 

In Section 3 we have shown how the nuclear 
matter equation of state can be calculated using in-medium 
chiral perturbation theory up to three loop order in the energy density, 
expanded in powers of the Fermi momentum $k_f$ (modulo functions 
of $k_f/m_\pi$). The empirical saturation point, the nuclear matter 
incompressibility, and the asymmetry energy at saturation can be well 
reproduced at order $\mathcal{O}(k_f^5)$ in the chiral expansion with just one
single high-momentum cutoff scale which parametrizes short-distance physics.
In this approach the nuclear matter saturation mechanism is entirely 
determined by the interplay between attraction from two-pion exchange processes 
and the stabilizing effects of the Pauli principle. 

The density dependence of the strength parameters $G^{(\pi)}(\rho)$ 
originating from pionic (chiral) fluctuations is calculated 
by equating the isoscalar-scalar, the isoscalar-vector, the 
isovector-scalar, and the isovector-vector 
self-energies (\ref{selfnm1}-\ref{SE-VS}) 
in the single-nucleon Dirac equation (\ref{Dirac-NM}) with 
those calculated using in-medium chiral perturbation theory: 
\begin{eqnarray} 
G_S^{(\pi)} \rho_s & = & \Sigma_S^{\rm CHPT}(k_f,\rho) \label{GS} \; ,\quad\quad\\ 
\label{49}
G_V^{(\pi)} \rho + \Sigma_r^{(\pi)}
   & = & \Sigma_V^{\rm CHPT}(k_f,\rho) \label{GV} \; ,\\ 
G_{TS}^{(\pi)} \rho_{s3} & = & \Sigma_{TS}^{\rm CHPT}(k_f,\rho) \label{GTS} \; ,\\ 
G_{TV}^{(\pi)} \rho_{3} & = & \Sigma_{TV}^{\rm CHPT}(k_f,\rho) \label{GTV} \; , 
\end{eqnarray} 
with
\begin{equation}
\Sigma^{(\pi)}_r = \frac{1}{2} \frac{\partial G_S^{(\pi)}}{\partial \rho}\rho_s^2 
   + \frac{1}{2} \frac{\partial G_V^{(\pi)}}{\partial \rho} \rho^2 + 
   \frac{1}{2} \frac{\partial G_{TS}^{(\pi)}}{\partial \rho}\rho_{s3}^2 
   + \frac{1}{2} \frac{\partial G_{TV}^{(\pi)}}{\partial \rho} \rho_{3}^2 \; .
\end{equation} 
The equations of state of symmetric and asymmetric nuclear matter calculated in CHPT 
give, via the Hugenholtz - van Hove theorem, the sums 
$U_{(T)}(k_f,k_f) = \Sigma_{(T)S}^{\rm CHPT}(k_f,\rho)+ 
\Sigma_{(T)V}^{\rm CHPT} 
(k_f,\rho)$ 
of the scalar and vector nucleon self-energies   
in the isoscalar and isovector channels 
at the Fermi surface $p=k_f$ up to two-loop order, generated by 
chiral one- and two-pion exchange \cite{KFW2}. 
The differences $\Sigma_{(T)S}^{\rm CHPT}(k_f,\rho)- 
\Sigma_{(T)V}^{\rm CHPT} (k_f,\rho)$ are 
calculated from the same pion-exchange diagrams via charge conjugation (formally evaluating 
the anti-nucleon single particle potential in nuclear matter). 
Following a procedure similar to the determination of the nucleon-meson vertices of relativistic mean-field models 
from Dirac-Brueckner calculations \cite{HKL.01}, we neglect the momentum 
dependence of $\Sigma_{(T)S,V}^{\rm CHPT}(p,\rho)$ and take their values at the 
Fermi surface $p=k_f$. A polynomial fit up to order $k_f^5$ is performed, 
and all four CHPT self-energies have the same functional form, 
\begin{eqnarray} 
   \Sigma^{CHPT}(k_f, \lambda) & = & \left[ c_{30} + c_{31}\lambda + 
   c_{32}\lambda^2 + c_{3L} \ln \frac{m_{\pi}}{4\pi f_{\pi}\lambda} \right] 
   \frac{k_f^3}{M^2}\nonumber\\ 
& ~ & + c_{40} ~ \frac{k_f^4}{M^3} + \left[ c_{50} + c_{5L} 
   \ln \frac{m_{\pi}}{4\pi f_{\pi}\lambda} \right] \frac{k_f^5}{M^4} \; , 
\label{expansion} 
\end{eqnarray} 
where the dimensionless parameter $\lambda$ is related to the 
momentum cut-off scale by 
$\Lambda = 2\pi f_{\pi} \lambda$, 
and $f_\pi =$ 92.4 MeV denotes the pion decay constant. 
The cut-off parameter $\Lambda \simeq 0.65$ GeV that has been adjusted to the 
nuclear matter saturation in Ref.\cite{KFW1}, corresponds to 
$\lambda = 1.113$. In order to determine the density dependence 
of the couplings from Eqs. (\ref{GS}-\ref{GTV}), the CHPT 
self-energies are re-expressed in terms of the baryon 
density $\rho = 2k_f^3/3\pi^2$: 
\begin{eqnarray} 
  \Sigma_S^{\rm CHPT}(k_f,\rho) & = & (c_{s1} + c_{s2}\rho^{\frac{1}{3}} 
  + c_{s3}\rho^{\frac{2}{3}}) \rho \; ,\label{prho1} \\ 
  \Sigma_V^{\rm CHPT}(k_f,\rho) & = & (c_{v1} + c_{v2}\rho^{\frac{1}{3}} 
  + c_{v3}\rho^{\frac{2}{3}}) \rho \; ,\label{prho2}\\ 
  \Sigma_{TS}^{\rm CHPT}(k_f,\rho) & = & (c_{ts1} + c_{ts2}\rho^{\frac{1}{3}} 
  + c_{ts3}\rho^{\frac{2}{3}}) \rho_3 \; ,\label{prho3}\\ 
  \Sigma_{TV}^{\rm CHPT}(k_f,\rho) & = & (c_{tv1} + c_{tv2}\rho^{\frac{1}{3}} 
  + c_{tv3}\rho^{\frac{2}{3}}) \rho_3 \label{prho4}\; . 
\end{eqnarray} 
The resulting expressions for the density dependent couplings 
of the pionic fluctuation terms are\footnote{
small differences between $\rho_s$ and $\rho$ at nuclear matter densities
are neglected here
} 
\begin{eqnarray} 
  G_S^{(\pi)} & = & c_{s1} + c_{s2} \rho^{\frac{1}{3}} 
  + c_{s3} \rho^{\frac{2}{3}} \; ,\\ 
\label{58}
  G_V^{(\pi)} & = & \bar{c}_{v1} + \bar{c}_{v2} \rho^{\frac{1}{3}} 
  + \bar{c}_{v3} \rho^{\frac{2}{3}}
  \quad \left\{ \begin{array}{ccl}
\bar{c}_{v1} & = & c_{v1} \\
\bar{c}_{v2} & = & (6c_{v2}-c_{s2} - \delta^2(c_{ts2} +c_{tv2}))/7 \\
\bar{c}_{v3} & = & (3c_{v3}-c_{s3} - \delta^2(c_{ts3} +c_{tv3}))/4 
\end{array} \right. \; ,\\
%
%
  G_{TS}^{(\pi)} & = & c_{ts1} + c_{ts2} \rho^{\frac{1}{3}} 
  + c_{ts3} \rho^{\frac{2}{3}} \; ,\\ 
  G_{TV}^{(\pi)} & = & c_{tv1} + c_{tv2} \rho^{\frac{1}{3}} 
  + c_{tv3} \rho^{\frac{2}{3}} \; , 
\end{eqnarray} 
where $\delta = (\rho_p - \rho_n)/(\rho_p + \rho_n)$.
For $\lambda =$ 1.113, the coefficients 
of the expansion of the CHPT self-energies in powers 
of the baryon density Eqs.(\ref{prho1}-\ref{prho4}) are given 
in Table \ref{tab2}. 


\begin{table} 
\begin{center} 
\caption{The coefficients of the expansion of the in-medium 
CHPT self energies (\protect\ref{prho1}) -- (\protect\ref{prho4}) 
in powers of the baryon density, for the value of the cut-off 
parameter $\Lambda =$ 646.3 MeV. In this case we assumed
$N=Z$, neglecting isovector contributions.} 
\bigskip 
\begin{tabular}{ccc|ccc} 
\hline 
& & & & &\\
c$_{s1}$ & -2.805 fm$^{2}$ & & & c$_{ts1}$ & -0.373 fm$^{2}$\\ 
c$_{s2}$ & 2.738  fm$^{3}$ & & & c$_{ts2}$ & 0.251 fm$^{3}$\\ 
c$_{s3}$ & 1.346  fm$^{4}$ & & & c$_{ts3}$ & 0.387 fm$^{4}$\\ 
c$_{v1}$ & -2.718 fm$^{2}$ & & & c$_{tv1}$ & -0.562 fm$^{2}$\\ 
c$_{v2}$ & 2.841  fm$^{3}$ & & & c$_{tv2}$ & 0.207 fm$^{3}$\\ 
c$_{v3}$ & 1.325  fm$^{4}$ & & & c$_{tv3}$ & 0.398 fm$^{4}$\\ 
& & & & &\\
\hline 
\end{tabular} 
\label{tab2} 
\end{center} 
\end{table} 

Combining now the effects from leading order QCD condensates and 
pionic fluctuations, 
the strength parameters of the isoscalar four-fermion 
interaction terms in the Lagrangian (\ref{Lag}) are: 
\begin{equation}
G_{S,V}(\rho) = G^{(0)}_{S,V} +  G^{(\pi)}_{S,V}(\rho)\,\,. 
\label{couplings} 
\end{equation} 
For the isovector channels we assume that only pionic 
(chiral) fluctuations contribute. 

\subsubsection{C. Corrections of higher order} 
 
Up to this point the nucleon self-energies are evaluated
at chiral two-loop level including terms of order $k_f^5$. At order
$k_f^6$ ($\propto \rho^2$), several additional effects appear.
First, condensates of higher dimension enter, such as
$\langle\bar{q}\Gamma q \bar{q} \Gamma q \rangle$,
$\langle q^\dagger q\rangle^2$. Their detailed
in-medium dependence is difficult to estimate.
Secondly, four-loop CHPT contributions to the energy density
(three-loop in the self-energies) introduce genuine
3-body interactions.
All those effects combine to produce the $\mathcal{O}(k_f^6)$ correction
in the energy per particle.
Our conjecture is that these terms should be subleading in
the $k_f$-expansion.
In estimating those contributions, we will take
a pragmatic point of view.
Given the density-independent leading condensate terms $G^{(0)}$ 
and the CHPT couplings $G^{(\pi)}(\rho)$ evaluated to 
$\mathcal{O} (k_f^5) \propto \rho^{\frac{2}{3}}$, we generalize
\begin{equation}
G(\rho) = G^{(0)} + G^{(\pi)} (\rho) + \delta G^{(1)} (\rho) \; ,
\label{higherorder}
\end{equation} 
adding terms $\delta G^{(1)} = g^{(1)} \rho$ and let the constant $g^{(1)}$ be
determined by a least squares fit to properties of nuclear matter and 
finite nuclei. At first sight, the corrections $\delta G^{(1)}_S,~
\delta G^{(1)}_V,{\rm etc}\ldots$ would then introduce several additional parameters,
undermining our startegy to keep the number of freely
adjustable fine-tuning constants as small as possible. For the scalar self-energy
a primary uncertainty arises from the four-quark condensate. It is common
practice to approximate $\langle \bar{q}q \bar{q}q \rangle$ assuming
factorization into the form $\langle \bar{q} q \rangle^2$, which introduces 
potentially large and uncontrolled errors  
as discussed in detail in Ref.~\cite{Jin.94}. While the in-medium values
of condensates such as $\langle\bar{q}\Gamma q \bar{q} \Gamma q \rangle$
are thus not well determined, it nevertheless appears that only a weak
or moderate density dependence of these condensates is consistent with known 
nuclear phenomenology.
We do not have to be much concerned about this issue here because our  
{\it explicit} treatment of scalar $\pi\pi$ fluctuations
(up to order $k_f^5$ in the energy density) removes at least part of such
uncertainties. What we can conclude from this calculation is that
factorization of the four-quark condensates, assuming ground state
dominanance, is not justified since the QCD ground state strongly couples
to the scalar two-pion continuum via the operator $\bar{q}q$.

So as a first option, 
we assume that the only non-vanishing $\delta G^{(1)}$
is $\delta G^{(1)}_V = g^{(1)}_V \rho$, and determine the 
constant $g^{(1)}_V$. Note that this $\delta G^{(1)}$ produces a self-energy
contribution $\delta \Sigma^{(1)} \propto \rho^2$ which acts like a three-body force.

\subsection{Nuclear matter, part III}

\index{nuclear matter equation of state (EOS)}
We return now to the calculation of the nuclear matter equation of state
(EOS), using the relativistic point-coupling model with QCD-constrained input
as outlined in the previous sections. These calculations are performed in
three steps. 

i) In the first step we follow the implicit assumption 
made in Ref.\cite{KFW1} and in Section 3, namely 
that $\Sigma_S^{(0)} = - \Sigma_V^{(0)}$ 
in nuclear matter, and ignore the contribution of the 
condensate background self-energies.  
The point-coupling model, with the density dependence of the strength 
parameters determined 
by pionic (chiral) fluctuations, nicely reproduces the 
nuclear matter EOS calculated previously using in-medium CHPT:
the binding energy per particle, the saturation density, the 
compression modulus, and the asymmetry energy. Small 
differences arise mainly because in the mapping of the 
CHPT nucleon self-energies on the self-energies of the 
point coupling model in Eqs. (\ref{GS}-\ref{GTV}), the 
momentum dependence of the former has been frozen to the 
values at the Fermi momentum. This is a well known problem that arises 
also in the determination of the meson-nucleon in-medium vertices 
from Dirac-Brueckner calculations of nucleon self-energies 
in nuclear matter~\cite{HKL.01}. 

At very low-densities, the CHPT equation of state cannot reproduce
realistic calculations of $E/A$, for obvious reasons.
The low density limit of the energy per particle is not given by a free
Fermi gas of nucleons. Proton-neutron pairs coalesce to form
deuteron bound-states, an entirely non-perturbative phenomenon
that cannot be handled at any order of a perturbative approach.
The low-density limit of $E/A$ approaches a constant
as $k_f \rightarrow 0$, namely one half of the deuteron binding energy.
The effect of this shift is marginal, however, for the bulk 
of nuclear matter around saturation density.

ii) The next step  
includes the contributions of the condensate background self-energies, $\Sigma_{S,V}^{(0)}$, 
in the isoscalar-scalar and isoscalar-vector channels. 
In Ref.\cite{FKVW1} we have shown that, even though to 
first approximation the condensate potentials do not play a role 
in the saturation mechanism, they are indeed essential for the 
description of ground-state properties of finite nuclei. 

At this stage we first treat both $G_S^{(0)}$ and $G_V^{(0)}$ as independent 
parameters, irrespective of the QCD sum rule constraint (\ref{ratio}).
The three parameters: $G_S^{(0)}$, $G_V^{(0)}$ and $\Lambda$ (the CHPT cutoff
scale), are then adjusted 
to reproduce ``empirical" nuclear matter properties: 
$E/A = -16$ MeV (5\%), $\rho_0 = 0.153$ fm$^{-3}$  (10\%), 
the compression modulus $K_0 = 250$ MeV (10\%), and the asymmetry energy
$A_0 = 33$ MeV (10\%). The values in   
parentheses refer to the error bars used in the fitting procedure. 

\begin{figure} 
\begin{center}
\includegraphics[scale=0.5,angle=-90]{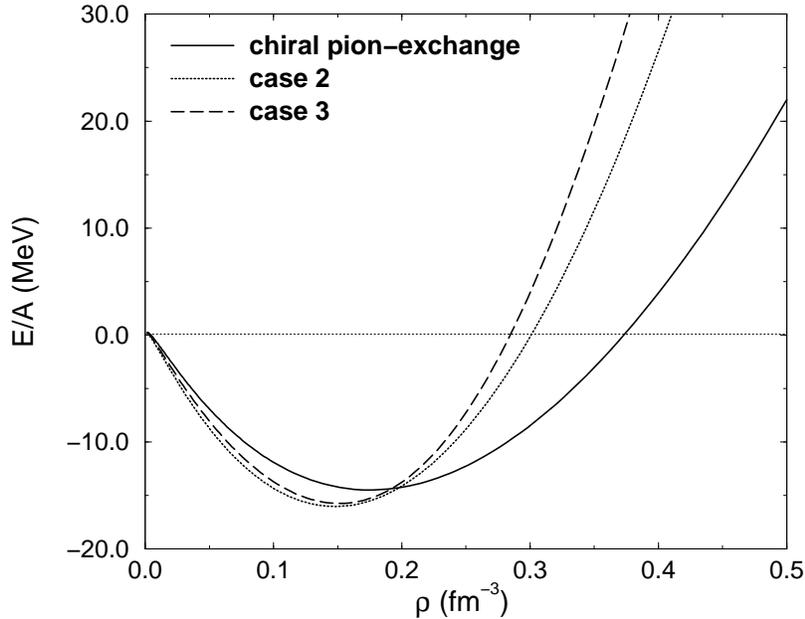} 
\end{center}
\caption{\label{figC} Binding energy per nucleon for symmetric nuclear 
matter as a function of the baryon density, calculated from 
chiral one- and two-pion exchange between nucleons (case 1, solid curve), 
by adding the isoscalar condensate background nucleon self-energies linear in the 
corresponding densities, with $G_V^{(0)} = - G_S^{(0)} = 7~{\rm fm}^2$ 
(case 2, dotted curve), and finally by including also the non-linear contribution 
$g_V^{(1)} \rho^2$ to the isoscalar vector condensate self-energy,
with $G_V^{(0)} = 11~{\rm fm}^2$, $G_S^{(0)} = -12~{\rm fm}^2$ 
(case 3, dashed curve).} 
\end{figure}

The corresponding {\it least-squares fit} yields 
$G_S^{(0)} =  -7$ fm$^2$, $G_V^{(0)} =  7$ fm$^2$, and $\Lambda = 685$ MeV.
The result is shown in 
Fig. \ref{figC} (EOS of isospin symmetric nuclear matter in 
comparison with the one calculated exclusively from chiral pion-exchange) 
and in the first row of Table \ref{tab4}. The EOS so obtained
is not yet satisfactory. The asymmetry energy $A_0$ at saturation is 
too high, and the values of $G_S^{(0)}$ and $G_V^{(0)}$ 
are smaller than the leading order QCD sum rule 
estimate (\ref{estimate}). This results in the relatively large Dirac 
effective mass $M^*/M =$ 0.75. A high Dirac mass in turn indicates 
that the effective single-nucleon spin-orbit potential is too 
weak to reproduce the empirical energy spacings between 
spin-orbit partner states in finite nuclei. 
Stronger background scalar and vector fields are required in order   
to drive the large spin-orbit splittings in finite nuclei. 

Notice however that, even though 
$G_S^{(0)}$ and $G_V^{(0)}$ were varied {\it independently}, the 
minimization procedure chooses to balance the contributions from the 
corresponding large scalar and vector self-energies, such that their 
sum tends to vanish: 
there is already sufficient binding from pionic fluctuations alone, 
so that even the unconstrained fit prefers $\Sigma_S^{(0)} = 
- \Sigma_V^{(0)}$ for the condensate potentials. 
This remarkable feature prevails as we now proceed 
with fine-tuning improvements. 
\begin{table}[t] 
\begin{center} 
\caption{Nuclear matter saturation properties calculated 
in the relativistic point-coupling model constrained by in-medium 
QCD sum rules and chiral perturbation theory. In addition to the 
chiral one- and two-pion exchange contribution to 
the density dependence of the coupling parameters, the 
nuclear matter EOS shown in the first row incorporates the 
isoscalar condensate background nucleon self-energies linear in the 
corresponding densities. The EOS displayed in the second row 
is calculated by including also the non-linear ``3-body'' contribution 
(proportional to $\rho^2$) in the isoscalar vector condensate self-energy.} 
\bigskip 
\begin{tabular}{cccccc} 
\hline 
\\
 ~ & E/A[MeV] & $\rho_0${\rm[fm$^{-3}$]} & $K_0$[MeV] & 
 $M^*/M$ & $A_0$[MeV]\\ 
 \\
\hline
\\ 
 {\sc linear} & -15.97 & 0.148 & 283 & 0.753 & 45.3 \\ 
 {\sc non-linear} & -15.76 & 0.151 & 332 & 0.620 & 30.2 \\ 
 \\
\hline 
\end{tabular} 
\label{tab4} 
\end{center} 
\end{table} 

iii) The third step incorporates in addition the higher-order 
correction $\delta G_V^{(1)} = g_V^{(1)} \rho$, see Eq.(\ref{higherorder}).
For $G_S^{(0)} = -12$ fm$^2$, 
$G_{V}^{(0)} =  11$ fm$^2$, $g_{V}^{(1)} =  -3.9$ fm$^5$ and 
$\Lambda = 600$ MeV, determined in a {\it least-squares fit} 
to the "empirical" nuclear matter input, 
the EOS resulting is displayed in Fig.~\ref{figC} 
and the nuclear matter properties at saturation are listed 
in the second row of Table \ref{tab4}. This nuclear matter EOS 
is actually quite satisfactory, with realistic values of the 
binding energy, saturation density, asymmetry energy at 
saturation, and a low effective Dirac mass. The only exception 
is a relatively large nuclear matter incompressibility. It could be 
cured by introducing a $\rho^3$ term in the expansion of the self-energies,
but we prefer not to invoke additional fine-tuning parameters at this
level of the discussion. 

The resulting couplings $G_{S,V}^{(0)}$ of the condensate background 
fields are remarkably close to the prediction (\ref{estimate}) of the 
leading order in-medium QCD sum rules. 
The scalar and the vector ``condensate''
self-energies $\Sigma_S^{(0)}=G_S^{(0)}\rho_s$ 
and $\Sigma_V^{(0)}=G_V^{(0)}\rho$ follow the expectation 
$\Sigma_S^{(0)} \simeq -\Sigma_V^{(0)}$ to within 
5\% at saturation density, even without this condition 
being pre-imposed.   
The large isoscalar condensate background self-energies 
in turn lead to a relatively low 
effective Dirac mass, crucial for the empirical spin-orbit 
splittings in finite nuclei. Finally, the cut-off   
$\Lambda = 600$ MeV differs by less than 10 \% from the value 
(646 MeV) obtained when the nuclear matter EOS results 
exclusively from one- and two-pion exchange between 
nucleons~\cite{KFW1}. 
The $\delta G^{(1)}_V = g^{(1)}_V\rho$ term is relatively small: at saturation density, $\delta G^{(1)}_V/G^{(0)}_V \simeq 0.05$.
Splitting $\delta G^{(1)}$ between scalar and vector parts
(e.g. by choosing $\delta G^{(1)}_V \simeq \delta G^{(1)}_S 
\simeq -2~{\rm fm}^{-5}$) would induce only marginal changes. 

\subsection{Comparison with Dirac-Brueckner G-matrix theory}

\index{Dirac-Brueckner G-matrix}
At this point it is instructive to compare
the density-dependence of our self-energies $\Sigma_{S,V} (\rho)$
with that of the scalar and the vector self-energies resulting from
Dirac-Brueckner G-matrix calculations which start from realistic NN-interactions $V_{NN}$.
Brueckner calculations iterate $V_{NN}$ to all orders in the ladder approximation, 
keeping track of Pauli exclusion priciple 
effects on intermediate $NN$ states. In addition, the Dirac-Brueckner approach takes into
account the lower components of the relativistic (Dirac) wave functions of the nucleons, so that the
density dependent nucleon self-energies resulting from such calculations separate into Lorentz scalar 
and vector parts. 
\begin{figure} 
\begin{center}
\vspace{0.75cm}
\includegraphics[scale=0.5]{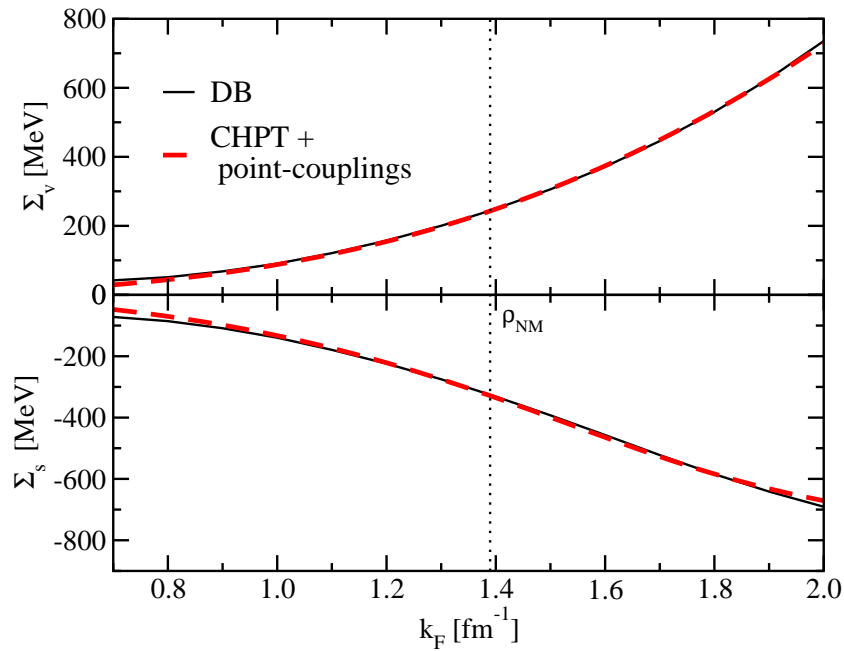} 
\end{center}
\caption{\label{fig.DB} Comparison of the $k_f$-dependencies of isoscalar 
vector and scalar self-energies resulting from a Dirac-Brueckner calculation
(solid lines: DB \cite{Gro.98}) with the self-energies generated from
in-medium chiral perturbation theory (dashed lines: CHPT + point-couplings) up
to 3-loop order in the energy density. (We thank Paolo Finelli for preparing
this figure.)}
\end{figure} 

We refer here explicitly to a calculation, reported in Ref. \cite{Gro.98},
which uses the Bonn A potential. With this potential,
the nuclear matter saturation density comes out somewhat too high ($\rho_{sat}
\simeq 0.185~{\rm fm}^{-3}$), higher
than the one of our best fit. In order to perform a meaningful comparison, simulating this
larger saturation density in our approach requires weakening the 
condensate mean fields $\Sigma_{S,V}^{(0)}$ by about $25$ \%, but with no changes
in the pionic terms $\Sigma_{S,V}^{(\pi)}$. After this readjustment
the difference in the $k_f$-dependences between our $\Sigma_S$ and
$\Sigma_V$ and those resulting from the Dirac-Bruckner calculation
is less than $10$ \% over the entire range of densities from $0.5\rho_0$ 
to $2.5\rho_0$ (see Fig. \ref{fig.DB}).
This is a remarkable observation:
it appears that in-medium chiral perturbation theory at two-loop order,
with a cut-off scale $\Lambda \simeq 0.6$ GeV that converts
unresolved short-distance dynamics at momenta beyond $\Lambda$ effectively
into contact terms, generates quantitatively similar in-medium nucleon self energies
as a full Dirac-Brueckner calculation, when requiring that both approaches
reproduce the same nuclear matter saturation point. The reasoning behind this 
observation can presumably be traced back to the separation of scales at work, a key 
element of chiral effective field theory  on which our approach is based.
One- and two-pion exchange in-medium dynamics at momentum scales
comparable to $k_f$ must be treated explicitly. 
Our in-medium CHPT approach includes all terms (ladders and others)
to three-loop order in the energy density, a prominent one being
iterated one-pion exchange. The non-trivial $k_f$
dependence in the self-energies (beyond ``trivial'' order $k_f^3$)
reflects the action of the Pauli principle in these processes. 
Such features are
also present in leading orders of the Brueckner
ladder summation which includes, for example, Pauli effects on 
iterated one-pion exchange.
On the other hand, the iteration of short-distance interactions to all orders
involves intermediate momenta much larger than $k_f$ and generates effectively
the same self-energy pieces (proportional to density $\rho$) as a suitably chosen contact 
interaction: the infinite ladder summation of short-distance interactions
can simply be subsumed in a contact term
reflecting the high-momentum scale $\Lambda$. The reasoning here has a close 
correspondence to related studies in Ref. \cite{Bog.03}.

\subsection{Intermediate summary and discussion}  

While the quantitative details for the terms contributing to the scalar
and vector self-energies can be reconstructed from the tables
and figures that we have included in this section, it is instructive
to observe systematic trends in these self-energies when expanded 
in powers of $k_f$ (or equivalently, in powers of $\rho^{\frac{1}{3}}$).
We focus on symmetric nuclear matter and write
\begin{equation}
\Sigma_{S,V} = \Sigma_{S,V}^{(0)} + \Sigma_{S,V}^{(\pi)}
+\delta\Sigma_{S,V} \; ,
\end{equation}
combining the leading condensate terms $\Sigma_{S,V}^{(0)}$,
the chiral (pionic) terms $\Sigma_{S,V}^{(\pi)}$ calculated
to order $k_f^5$, and possible corrections of higher order
summarized in $\delta \Sigma_{S,V}$.
We introduce the saturation density $\rho_0 = 0.16~{\rm fm}^{-3}$
as a convenient scale and write the condensate terms as
\begin{eqnarray}
\Sigma_S^{(0)} (\rho) & \simeq & - 0.35\,{\rm GeV}\, \frac{|G_S^{(0)}|}{11\,{\rm fm}^2}
\left( \frac{\rho_s}{\rho_0} \right) \; ,\\
\Sigma_V^{(0)} (\rho) & \simeq & + 0.35\,{\rm GeV}\, \frac{|G_V^{(0)}|}{11\,{\rm fm}^2}
\left( \frac{\rho}{\rho_0} \right) \; .
\end{eqnarray}
We note that  with $|G_S^{(0)}|$ chosen less than 5\% larger than $G_V^{(0)}$,
this just compensates for the difference between the baryon density $\rho$
and the scalar density $\rho_s < \rho$ such that $\Sigma_S^{(0)}
\simeq - \Sigma_V^{(0)}$ results at $\rho = \rho_0$.
The ``best-fit'' values $G_S^{(0)} \simeq -12\,{\rm fm}^2$ and
 $G_V^{(0)} \simeq + 11\,{\rm fm}^2$ are thus perfectly consistent
with QCD sum rule expectations: a non-trivial result.
The pionic fluctuations from one- and two-pion exchange processes
have the following approximate pattern:
\begin{equation}
\Sigma_{S,V}^{(\pi)} (\rho) \simeq - 75\,{\rm MeV}\,\left(\frac{\rho}{\rho_0} \right)
\left[ 1 + d_1 \left(\frac{\rho}{\rho_0} \right)^{\frac{1}{3}} +
 d_2 \left(\frac{\rho}{\rho_0} \right)^{\frac{2}{3}}\right] \; ,
\end{equation}
where $d_1$ varies between $-0.61$ (scalar) and $-0.65$ (vector) and
$d_2 \simeq -0.17$ for both cases. Finally, the higher order corrections are summarized
as 
\begin{equation} 
 \delta \Sigma_S + \delta \Sigma_V \simeq -20\,{\rm MeV}~\left( \frac{\rho}{\rho_0}
\right)^2 \; ,
\end{equation}
where the detailed decomposition into scalar and vector part turns out not to
be relevant.

The leading attraction at $\mathcal{O}(k_f^3)$ in $\Sigma_{S,V}^{(\pi)}$
depends on the cut-off scale $\Lambda \equiv 2 \pi f_\pi \lambda$
which separates ``active'' two-pion exchange dynamics at long and intermediate
distances from 
short-distance (high momentum) dynamics. Details at short distance
scales (related to the intrinsic structure of the nucleons and their
short-range interactions) are not resolved as long as the Fermi momentum 
$k_f$ is small compared to the chiral symmetry breaking scale,
$4\pi f_\pi$. The unresolved short-distance information
can then be translated into an equivalent (density-independent)
four-point vertex which generates self-energy terms linear
in the density at the mean-field level.

\section{Finite Nuclei}

In this section the relativistic nuclear point-coupling model, 
constrained by in-medium QCD sum rules and chiral perturbation theory, 
is applied in calculations of ground state properties of finite nuclei. 
In the mean-field approximation 
the ground state of a nucleus with A nucleons is represented by 
the antisymmetrized product of the lowest occupied single-nucleon 
stationary solutions of the Dirac equation~(\ref{Dirac}). 
The calculation is self-consistent in the sense that the 
nucleon self-energies are functions of nucleon densities and currents, 
calculated from the solutions of Eq.~(\ref{Dirac}). 
The mean-field approach to nuclear structure represents an 
approximate implementation of Kohn-Sham 
\tindex{density functional theory (DFT)} \cite{KS.65,Kohn,DG.90}, 
which is successfully employed in the treatment of the quantum
many-body problem in atomic, molecular and condensed matter physics. 
At the basis of the DFT approach are energy functionals of the 
ground-state density. In relativistic mean-field models 
these become functionals of the ground-state scalar density and 
of the baryon current. The scalar and vector self-energies
play the role of local relativistic Kohn-Sham 
potentials~\cite{SW.97,FS.00}. The mean-field models approximate
the exact energy functional, which includes all higher-order 
correlations, by powers and gradients of fields or densities,
with the truncation determined by power counting~\cite{FS.00}.  

\subsection{Ground state energy}

In this lecture we only consider spherical even-even nuclei. Because of 
time-reversal invariance the spatial 
components of the four-currents vanish, and the nucleon self-energies 
(reduced to the time component $\Sigma^0$ of the vector terms 
$\Sigma^\mu$) read: 
\begin{eqnarray} 
\Sigma^{0} & = & G_V \,\rho + D_V \triangle \rho - 
   eA^0\frac{1+\tau_3}{2} \; ,\\ 
\Sigma^{0}_{T} & = & G_{TV} \,\rho_{3} + D_{TV} \triangle 
   \rho_{3} \; ,\\ 
\Sigma_S & = & G_S \,\rho_s + D_S \triangle \rho_s \; ,\\ 
\Sigma_{TS} & = & G_{TS} \,\rho_{s3} + D_{TS} \triangle \rho_{s3} \; ,\\ 
\Sigma_{rS} & = & \frac{\partial D_S}{\partial \rho} 
   (\nabla \rho_s ) \cdot(\nabla \rho ) \; ,\\ 
\Sigma_{rTS} & = & \frac{\partial D_{TS}}{\partial \rho} 
   (\nabla \rho_{s3} ) \cdot(\nabla \rho ) \; ,\\ 
\Sigma_{rT}^{0} & = &\frac{\partial D_{TV}}{\partial \rho} 
   (\nabla \rho_{3} ) \cdot (\nabla \rho ) \; ,\\ 
\Sigma_r^{0} & = & + \frac{1}{2} \frac{\partial G_S}{\partial \rho} 
   \rho_s^2 - \frac{1}{2} \frac{\partial D_S}{\partial \rho} 
   (\nabla \rho_s ) \cdot (\nabla \rho_s )\nonumber\\ 
 & ~ &  + \frac{1}{2} \frac{\partial G_V}{\partial \rho} \rho^2 
   - \frac{1}{2} \frac{\partial D_V}{\partial \rho} 
   (\nabla \rho ) \cdot (\nabla \rho ) \nonumber\\ 
 & ~ &  + \frac{1}{2} \frac{\partial G_{TS}}{\partial \rho} 
   \rho_{s3}^2 - \frac{1}{2} \frac{\partial D_{TS}}{\partial \rho} 
   (\nabla \rho_{s3} ) \cdot (\nabla \rho_{s3} )\nonumber\\ 
 & ~ &  + \frac{1}{2} \frac{\partial G_{TV}}{\partial \rho} 
   \rho_{3}^2 - \frac{1}{2} \frac{\partial D_{TV}}{\partial \rho} 
   (\nabla \rho_{3}) \cdot(\nabla \rho_{3})\nonumber\\ 
 & ~ &  + \frac{\partial D_V}{\partial \rho} (\nabla \rho ) 
   \cdot (\nabla \rho ) \; . 
\end{eqnarray} 
In the {\it mean-field} approximation the isoscalar-scalar, 
isoscalar-vector, isovector-scalar and isovector-vector ground state 
densities are calculated from the Dirac wave functions $\psi_k$ 
of the occupied positive-energy orbits as: 
\begin{eqnarray} 
\rho_s & = & \sum\limits_{k=1}^{A} \bar \psi_k \psi_k \; ,\\ 
\rho & = & \sum\limits_{k=1}^{A} \psi^{\dagger}_k \psi_k \; ,\\ 
\rho_{s3} & = & \sum\limits_{k=1}^{A} \bar \psi_k \tau_3 \psi_k \; ,\\ 
\rho_{3} & = & \sum\limits_{k=1}^{A} \psi^{\dagger}_k \tau_3 \psi_k \; , 
\end{eqnarray} 
respectively. 
The expression for the ground state 
energy of a nucleus with A nucleons reads: 
\begin{eqnarray} 
E & = & \sum\limits_{k=1}^{A} \epsilon_k - \frac{1}{2} \int d^3x \left\{ 
   ~G_S \,\rho_s^2 + D_S \,\rho_s \triangle \rho_s + 
   G_V \rho^2 + D_V \rho \triangle \rho \right. \nonumber\\ 
 & ~ & + G_{TV}\rho_{3}^2 + D_{TV} \rho_{3} \triangle \rho_{3}   
   + G_{TS} \,\rho_{s3}^2 + D_{TS} \,\rho_{s3}\, \triangle \rho_{s3} 
   + j^p V_C \nonumber\\ 
 & ~ & + \frac{\partial G_S}{\partial \rho} \rho_s^2 \,\rho + 
   \frac{\partial G_V}{\partial \rho} \rho^3 + 
   \frac{\partial G_{TV}}{\partial \rho} \rho_3^2 \,\rho + 
   \frac{\partial G_{TS}}{\partial \rho} \rho_{s3}^2\, \rho  \nonumber\\ 
 & ~ & - \frac{\partial D_S}{\partial \rho} (\nabla \rho_s) \cdot 
   (\nabla \rho_s) \rho + 
   \frac{\partial D_V}{\partial \rho} (\nabla \rho) \cdot 
   (\nabla \rho) \rho  \nonumber\\ 
 & ~ & - \frac{\partial D_{TV}}{\partial \rho} 
   (\nabla \rho_{3}) \cdot (\nabla \rho_{3}) \rho 
   - \frac{\partial D_{TS}}{\partial \rho} 
   (\nabla \rho_{s3}) \cdot (\nabla \rho_{s3}) \rho   \nonumber\\ 
 & ~ & + 2\frac{\partial D_S}{\partial \rho} (\nabla \rho_s) 
   \cdot (\nabla \rho) \rho_s + 2\frac{\partial D_{TV}}{\partial \rho} 
   (\nabla \rho_{3}) \cdot (\nabla \rho) \rho_{3} \nonumber\\ 
 & ~ & + \left. 2\frac{\partial D_{TS}}{\partial \rho} 
   (\nabla \rho_{s3}) \cdot (\nabla \rho) \rho_{s3} \right\} \; ,   
\end{eqnarray} 
where $\epsilon_k$ denotes the single-nucleon energies. 
In addition, for open shell nuclei pairing correlations are 
included in the simple BCS approximation~\cite{GRT.90}. 
After the solution of the self-consistent Dirac 
equation, the microscopic estimate for the center-of-mass correction 
\begin{equation} 
\label{cms} 
E_{cm} = - \frac{<P_{cm}^2>}{2AM} \;, 
\end{equation} 
is subtracted from the total binding energy. 
Here $P_{cm}$ is the total momentum of the 
nucleus with $A$ nucleons. 

\subsection{Surface (derivative) terms}

The calculated properties of finite nuclei depend of course 
on an accurate tuning of coupling parameters. We keep the variational freedom 
in those parameters at minimum, constraining the 
density-dependent couplings $G_S$, $G_V$, $G_{TS}$ and $G_{TV}$ 
by in-medium 
QCD sum rules and explicit chiral perturbation theory calculations of one- and 
two-pion exchange diagrams, as described in the previous sections.
For finite nuclei, fine-tuning of their detailed surface structure will be
required, so we must also 
determine the coupling parameters of the derivative terms: 
$D_S$, $D_V$, $D_{TS}$ and $D_{TV}$. 
Dimensional considerations suggest the following ansatz 
\begin{equation} 
D(\rho) = \frac{G(\rho)}{{\mathcal M}^2} \; , 
\end{equation} 
where ${\mathcal M}$ is a characteristic mass scale for a given 
spin-isospin channel. 
There is no deeper reason, however, for the 
derivative terms to have the same density dependence as the coupling 
parameters of the four-fermion interactions. The density dependence 
of $D_i(\rho)$ could, in principle, be derived by CHPT calculations 
for inhomogeneous nuclear matter. A simpler option, followed here, 
is to treat 
the $D$'s as density-independent adjustable constants. 
In this case the remaining rearrangement 
contribution to the vector-self energy becomes, of course, much simpler. 
As it has been emphasized by Serot and Furnstahl~\cite{FS.00b}, 
the empirical data set of bulk 
and single-particle properties of finite nuclei can only constrain 
six or seven parameters in the general expansion of the effective 
Lagrangian in powers of the fields and their derivatives. In particular, 
only one parameter of the derivative terms can be determined 
by the binding energies and radii of spherical nuclei. In the present 
analysis we therefore set $D_V$, $D_{TS}$ and $D_{TV}$ equal to zero, and 
adjust the single remaining surface parameter $D_S$ of the 
isoscalar-scalar derivative term 
to properties of light and medium-heavy $N \approx Z$ nuclei.
This approximation, which was first used by Serot and 
Walecka in Ref.~\cite{SW.79}, implies that the isoscalar-vector, 
the isovector-scalar and the isovector-vector interactions are 
considered to be purely contact interactions (no gradient terms).  
\begin{figure}[t] 
\begin{center}
\includegraphics[scale=0.5,angle=-90]{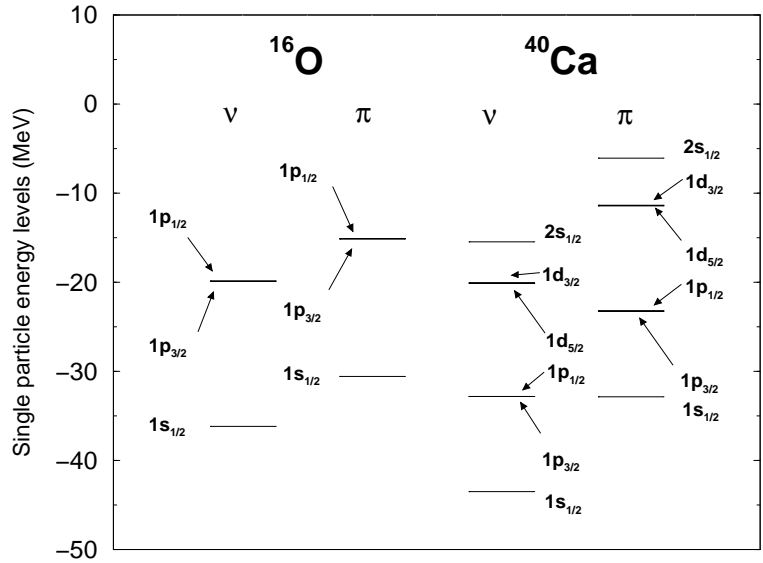} 
\end{center}
\caption{\label{figF} Neutron and proton single-particle levels 
in $^{16}$O and $^{40}$Ca calculated in the relativistic point-coupling model. 
The density dependent coupling strengths include only
the contribution from chiral one- and two-pion exchange between nucleons.
(From Ref.~\protect\cite{FKVW2}.)} 
\end{figure} 

\subsection{Single-particle energies}

It is instructive to consider separately the contributions from chiral 
pion dynamics and condensate background self-energies
to properties of finite nuclei. 
In a first step we have calculated the ground states of   
$^{16}$O and $^{40}$Ca using the coupling parameters determined 
by the nuclear matter EOS of Ref.~\cite{KFW1}: 
$G_S(\rho) =  G^{(\pi)}_S(\rho)$, 
$G_V(\rho) =  G^{(\pi)}_V(\rho)$, 
$G_{TS}(\rho) =  G^{(\pi)}_{TS}(\rho)$, 
$G_{TV}(\rho) =  G^{(\pi)}_{TV}(\rho)$, $\Lambda = 646.3$ MeV, 
while the couplings $G_{S,V}^{(0)}$
to the condensate background fields are set to zero. 
In this case the nuclear dynamics 
is completely determined by chiral (pionic) fluctuations. 
The resulting total binding energies 
are already within $5-8$ \% of the experimental values, but the radii 
of the two nuclei are too small (by about 0.2 fm). This is because 
the spin-orbit partners $(1p_{3/2},1p_{1/2})$ and 
$(1d_{5/2},1d_{3/2})$ are practically degenerate. In 
Fig.~\ref{figF} we display the calculated 
neutron and proton single-particle levels 
of $^{16}$O and $^{40}$Ca. The energies of the degenerate doublets 
are close to the empirical positions of the centroids of the 
spin-orbit partner levels, and even the calculated energies 
of the $s$-states are realistic. This is an interesting result. 
Chiral pion dynamics provides the saturation mechanism 
and binding of nuclear matter, but not the strong
spin-orbit force. The inclusion of the isoscalar-scalar 
derivative term has some effect on the calculated radii but 
it cannot remove the degeneracy of the spin-orbit doublets. 
\begin{figure} 
\begin{center}
\includegraphics[scale=0.5,angle=-90]{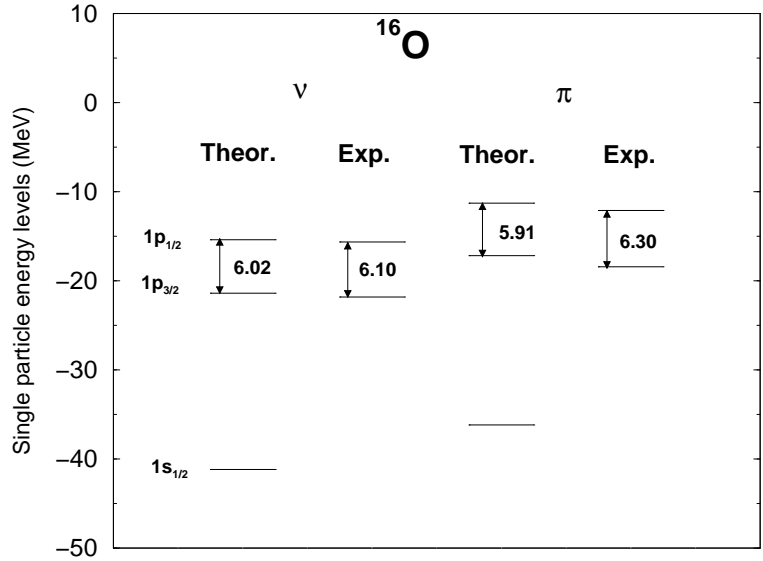}
\end{center} 
\caption{\label{figG} The neutron and proton single-particle levels 
in $^{16}$O calculated in the relativistic point-coupling model, are shown 
in comparison with experimental levels. The calculation is performed 
by including both the contributions of chiral pion-nucleon exchange and 
of the isoscalar condensate self-energies.(From Ref.~\protect\cite{FKVW2}.)} 
\end{figure} 

The \tindex{spin-orbit potential} plays a central role in 
nuclear structure. It is at the basis of the nuclear 
shell model, and its inclusion is essential in order to 
reproduce the experimentally established magic numbers. 
In non-relativistic models based on the mean-field approximation, 
the spin-orbit potential is included in a purely phenomenological way, 
introducing the strength of the spin-orbit interaction as 
an additional parameter. Its value 
is usually adjusted to the experimental 
spin-orbit splittings in spherical nuclei, for example $^{16}$O. 
On the other hand, in the 
relativistic description of the nuclear many-body problem, 
the spin-orbit interaction arises naturally from the 
scalar-vector Lorentz structure of the effective Lagrangian.
In the first order approximation, and assuming spherical 
symmetry, the spin-orbit term of the effective single-nucleon 
potential can be written as 
\begin{equation}
\label{so1}
  V_{s.o.} = \frac{1}{2M^2} \left(
  {1 \over r}{\partial \over \partial r} V_{ls}(r)\right) {\bm l \cdot \bm s} \; ,
\end{equation}
where the large spin-orbit potential $V_{ls}$ arises from the difference of the vector
and scalar potentials, $V-S \sim 0.7 ~{\rm GeV}$ \cite{Rin.96,Bro.77}. Explicitly, 
\begin{equation} 
\label{so2} 
V_{ls} = {M \over M_{eff}} (V-S) \; . 
\end{equation} 
where 
$M_{eff}$ is an effective mass specified as \cite{Rin.96}
\begin{equation} 
\label{so3} 
M_{eff} = M - {1 \over 2} (V-S). 
\end{equation} 
The isoscalar nucleon self-energies generated by pion exchange 
are not sufficently large to produce the empirical effective 
spin-orbit potential. 
The degeneracy of spin-orbit doublets is removed by 
including the isoscalar background self-energies $\Sigma_{S,V}^{(0)}$
that arise through changes in the quark condensate and the quark density 
at finite baryon density. 

\begin{figure} 
\begin{center}
\includegraphics[scale=0.5,angle=-90]{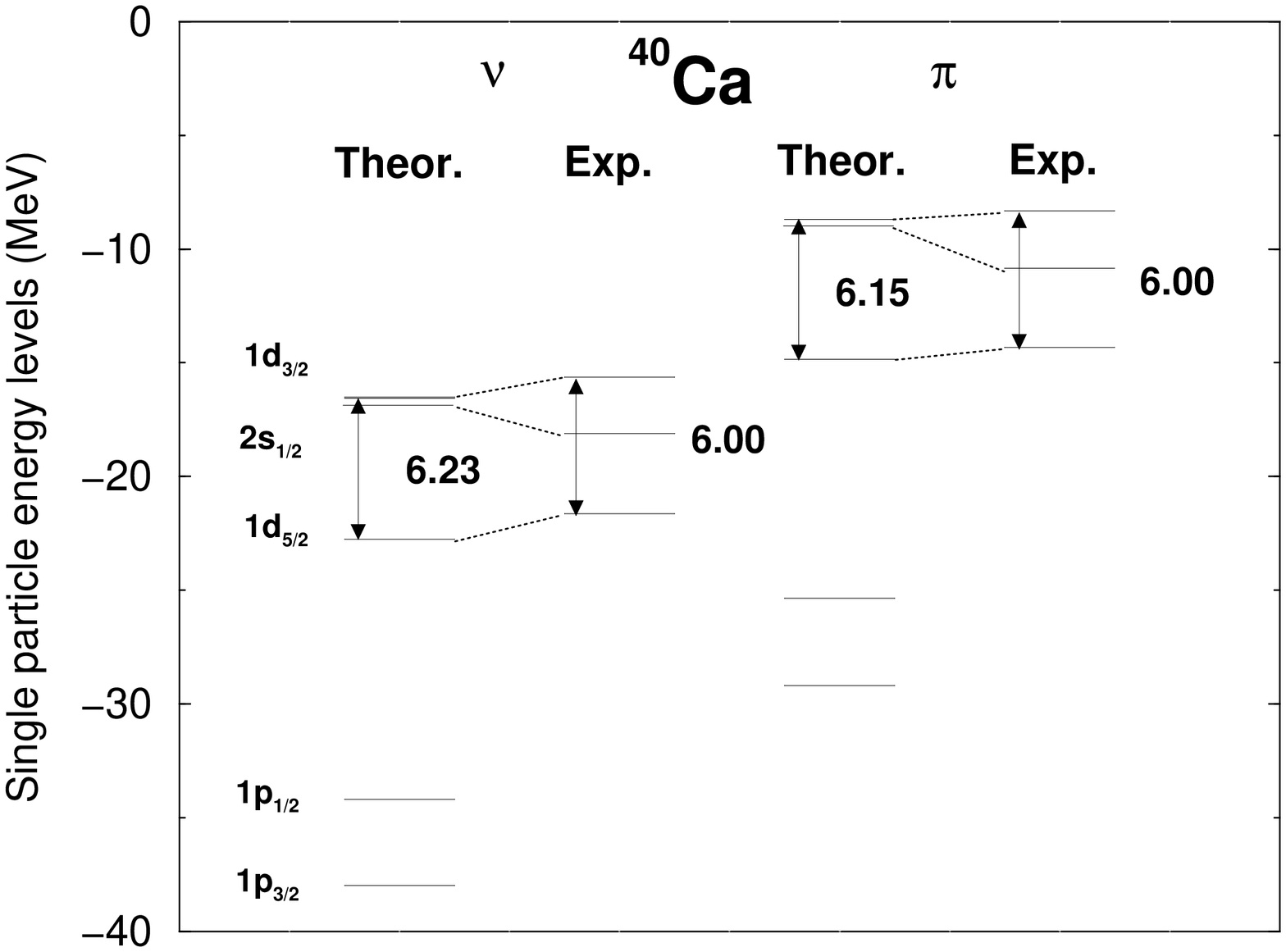} 
\end{center}
\caption{\label{figH} Same as in Fig.~\protect\ref{figG}, but for 
$^{40}$Ca.(From Ref.~\protect\cite{FKVW2}.)} 
\end{figure} 
\begin{figure} 
\begin{center}
\includegraphics[scale=0.5,angle=-90]{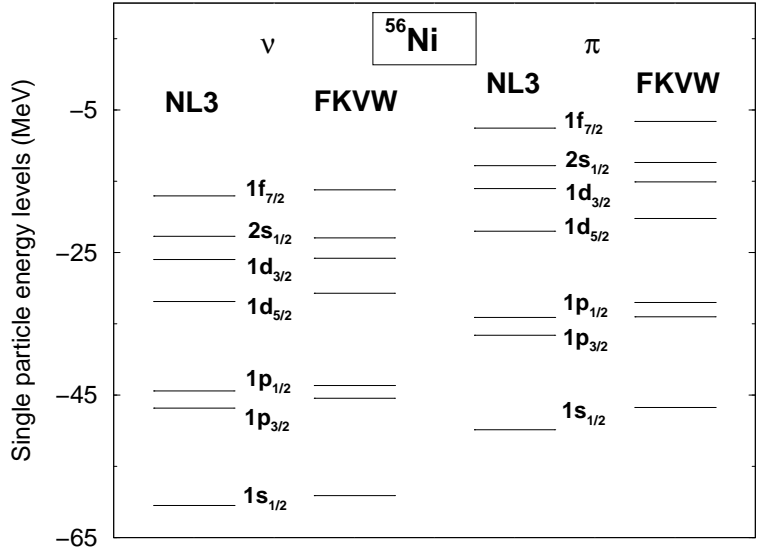} 
\end{center}
\caption{\label{figK}Neutron and proton single-particle spectra of 
$^{56}$Ni, calculated with the standard relativistic mean-field 
model NL3 effective interaction~\protect\cite{LKR.97}, and 
with the relativistic 
point-coupling model constrained by in-medium 
QCD sum rules and chiral perturbation theory (FKVW \protect\cite{FKVW2}).} 
\end{figure} 
The effect of including $\Sigma_{S,V}^{(0)}$ is demonstrated 
in Figs.~\ref{figG} and \ref{figH}. In addition to the four parameters
($G_S^{(0)} = -12$ fm$^2$, 
$G_{V}^{(0)} =  11$ fm$^2$, $g_{V}^{(1)} =  -3.9$ fm$^5$ and 
$\Lambda = 600$ MeV) 
determined by the nuclear matter equation 
of state, we have adjusted the isoscalar-scalar derivative term: 
$D_S = -0.713$ fm$^{4}$. 
This value of $D_S$ is very close to the ones used in the effective 
interactions of the standard relativistic point-coupling model 
of Ref.~\cite{Burvenich:2001rh}. 
It is also consistent with the ``natural'' order of magnitude expected
from $D_S \sim G_S/\Lambda^2$. 

The calculated neutron and proton single-particle 
energies of $^{16}$O and $^{40}$Ca are in excellent agreement 
with the empirical 
single-nucleon levels in the vicinity of the Fermi surface. 
In particular, with the inclusion of the 
isoscalar condensate self-energies, 
the model reproduces the empirical energy differences between 
spin-orbit partner states. This is an important result. It supports
our primary conjecture: while nuclear binding and 
saturation are almost completely generated by chiral (two-pion exchange) 
fluctuations in our approach, strong scalar and 
vector fields of equal magnitude and 
opposite sign, induced by changes of the QCD vacuum in the presence 
of baryonic matter, generate the large effective 
spin-orbit potential in finite nuclei. Not
surprisingly, the$~^{40}$Ca spectrum is reminiscent of an underlying 
pseudo-spin symmetry~\cite{Gin.02}.

In Fig. \ref{figK} we  compare the single-nucleon spectra 
of $^{56}$Ni, calculated in our approach,
with the results of a relativistic mean-field calculation using NL3~\cite{LKR.97},
probably the best phenomenological non-linear meson-exchange effective interaction.  
The agreement is convincing. All these results 
demonstrate that in the present approach, based on 
QCD sum rules and in-medium chiral perturbation theory, and 
with a small number of model parameters determined 
directly by these constraints, 
it is possible to describe symmetric and asymmetric nuclear matter, 
as well as properties of finite nuclei, at the same quantitative level as 
the best phenomenological relativistic mean-field models. 

\subsection{Systematics of binding energies and charge radii}

In Table~\ref{tab5} we summarize our calculated  
binding energies per nucleon and charge radii of light 
and medium-heavy nuclei, in comparison 
with experimental values. The parameters have been kept unchanged from those
used in the $^{16}$O and $^{40}$Ca calculations. The resulting agreement between 
the calculated and empirical binding energies and charge radii is indeed very good. 

\begin{table} 
\begin{center} 
\caption{Binding energies per nucleon and charge radii of light 
and medium-heavy nuclei, calculated in the relativistic 
point-coupling model constrained by in-medium 
QCD sum rules and chiral perturbation theory \cite{FKVW2}, are compared 
with experimental values.} 
\bigskip 
\begin{tabular}{ccccc} 
\hline 
 ~ & $E/A^{\rm exp}~({\rm MeV})$ & $E/A~({\rm MeV})$ & 
    $r_c^{\rm exp}~({\rm fm}^{-3})$ 
   & $r_c~({\rm fm}^{-3})$ \\ 
\hline 
 $~^{16}{\rm O}$  & 7.976 & 8.027 & 2.730 & 2.735 \\ 
 $~^{40}{\rm Ca}$ & 8.551 & 8.508 & 3.485 & 3.470 \\ 
 $~^{42}{\rm Ca}$ & 8.617 & 8.537 & 3.513 & 3.473 \\ 
 $~^{48}{\rm Ca}$ & 8.666 & 8.964 & 3.484 & 3.486 \\ 
 $~^{42}{\rm Ti}$ & 8.260 & 8.182 & ----- & 3.551 \\ 
 $~^{50}{\rm Ti}$ & 8.756 & 8.779 & ----- & 3.571 \\ 
 $~^{52}{\rm Cr}$ & 8.776 & 8.635 & 3.647 & 3.641 \\ 
 $~^{58}{\rm Ni}$ & 8.732 & 8.493 & 3.783 & 3.778 \\ 
 $~^{64}{\rm Ni}$ & 8.777 & 8.775 & 3.868 & 3.879 \\ 
 $~^{88}{\rm Sr}$ & 8.733 & 8.855 & 4.206 & 4.234 \\ 
 $~^{90}{\rm Zr}$ & 8.710 & 8.746 & 4.272 & 4.284 \\ 
\hline 
\end{tabular} 
\label{tab5} 
\end{center} 
\end{table} 

\subsection{Density distributions}

\begin{figure} 
\begin{center}
\includegraphics[scale=0.5,angle=-90]{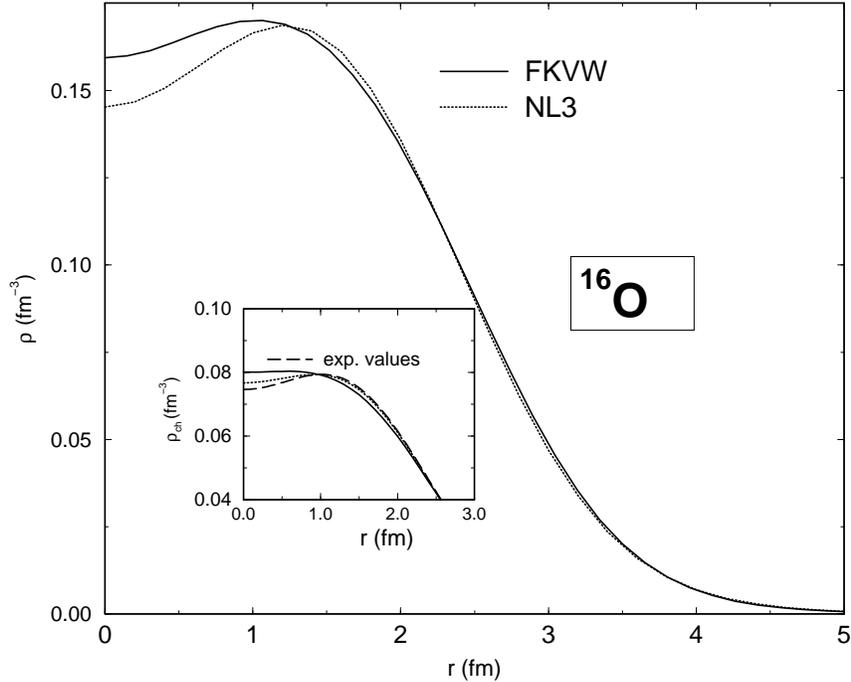} 
\end{center}
\caption{\label{figI} Nucleon density of $^{16}$O as a function 
of the radial coordinate. The result obtained in the QCD-constrained
relativistic point-coupling model (FKVW \protect\cite{FKVW2}) is compared with the 
nucleon density calculated in the standard relativistic mean-field 
model with the NL3 effective interaction~\protect\cite{LKR.97}. 
In the insert the corresponding charge densities are compared 
with the empirical charge density profile~\protect\cite{Vri.87}.} 
\end{figure}

\begin{figure}
\begin{center} 
\includegraphics[scale=0.5,angle=-90]{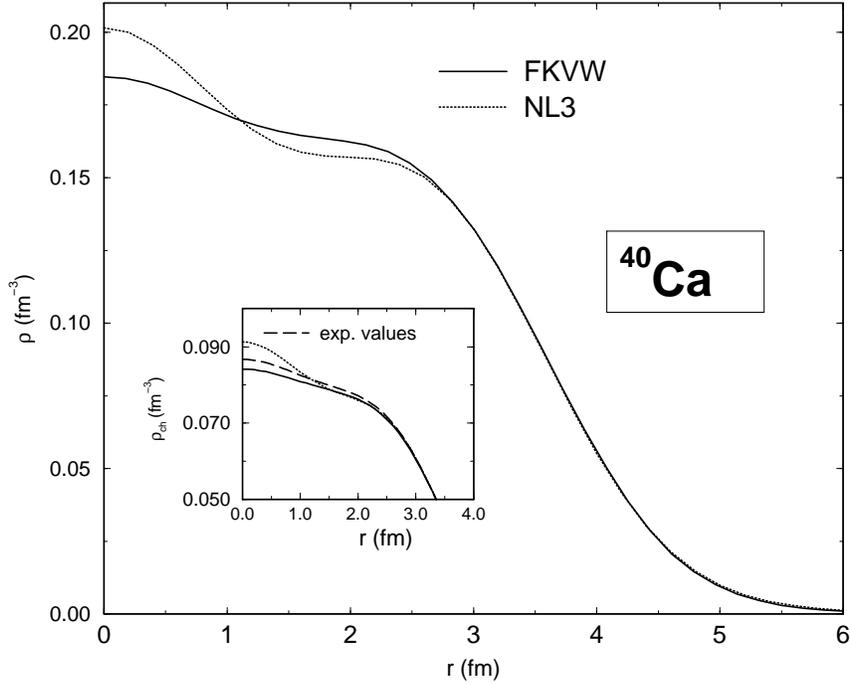} 
\end{center}
\caption{\label{figJ} Same as in Fig.~\protect\ref{figI}, but for 
$^{40}$Ca.} 
\end{figure}

Calculated nucleon densities of $^{16}$O and $^{40}$Ca 
are plotted in Figs. \ref{figI} and  \ref{figJ}. 
The results obtained in our QCD-constrained relativistic 
point-coupling model \cite{FKVW2} are compared with the 
nucleon densities calculated in the standard relativistic mean-field 
meson-exchange model with the NL3 effective interaction~\cite{LKR.97}. 
Also shown are the corresponding charge densities in comparison 
with the empirical charge density profiles~\cite{Vri.87}. 
When compared not only with NL3, but also with other standard 
relativistic mean-field effective interactions, the density 
profiles calculated in our point-coupling model display less 
pronounced shell effects, in better agreement with empirical 
densities. 

\section{Concluding remarks and outlook}

Our aim in these lectures notes has been to 
demonstrate the key role of a guiding principle, emerging from the
non-perturbative dynamics of QCD: spontaneous chiral symmetry breaking.
The effective field theory based on this principle ties together a wide
variety of strong-interaction physics involving the lightest quarks:
from the low-energy interactions of pions and nucleons via the condensate structure of
the QCD vacuum and its global thermodynamical implications, to 
fundamental aspects of nuclear binding and saturation. Much work lies ahead; the 
exploration of the nuclear many-body problem in view of its low-energy QCD connections
promises new insights. At the least, the QCD constraints can significantly
reduce the freedom in the choice of parameters. The wide variety of results reported 
in previous sections have been obtained with only five parameters, two of which (the
couplings $G_{S,V}^{(0)}$) turn out to be so remarkably close to leading order QCD sum rule
estimates that one could have guessed their values right from the beginning. The remaining
three parameters (the high-momentum scale reflecting unresolved short-distance physics,
a "three-body" contribution to the nucleon self-energy and a surface (derivative) term) behave
"naturally" according to the power-counting doctrine of effective field theory.

Of course, questions about
systematic convergence of the in-medium chiral loop expansion still remain and
need to be further explored. In particular, the important role of the $\Delta(1230)$, the prominent spin-isospin excitation of the 
nucleon, is so far just hidden in the parametrization of "short-distance" phenomena. However,
the scale associated with the mass difference $M_\Delta - M_N \sim 0.3$ GeV is in fact comparable
to the nuclear Fermi momentum, suggesting a more explicit treatment of the $\Delta$ in the chiral effective Lagrangian.

Calculations for heavier $N \neq Z$ nuclei are in progress.  Systems such as $^{208}$Pb
and beyond are a test case for the detailed isospin structure of the underlying chiral pion dynamics.
In the present version of the model, the isovector parts of the interaction  
are exclusively determined by chiral one- and 
two-pion exchange. Global properties of asymmetric nuclear matter and neutron matter
are reproduced quite well at least at moderate densities, but pion exchange processes do not account for 
the short-range part of the isovector effective interaction: the resulting density dependent couplings $G_{TS}$ and $G_{TV}$ are presumably still too weak. There is need for stronger self-energies 
in the isovector channel. In meson-exchange language, such additional short-distance interactions
could arise from the $\rho$ and $a_0$ resonances 
(where we must note, of course, that a large non-resonant 
low-mass part of the $\rho$ meson channel is already accounted for 
by explicit isovector two-pion exchange). This is 
an important topic, to be further addressed in future studies 
which will focus on a quantitative description of heavy nuclei and extrapolations
into regions of extreme isospin.

\subsubsection{Acknowledgements}

We would like to thank Norbert Kaiser whose work is at the basis of the developments
reported in these notes, and Paolo Finelli and Stefan Fritsch for their contributions to
this joint venture.

\end{document}